\documentclass[preprint,showpacs,aps,prc,groupedaddress,floatfix]{revtex4}
\usepackage{epsfig}
\usepackage{amsmath,amssymb} \usepackage{bm}

\newcommand{\vect}[1]{\mbox{${\bf #1}$}}

\newcommand{\pfd}[2]{{\frac{\partial #1}{\partial #2}}}

\newcommand{\vdot}{\mbox{\boldmath $\cdot$}}

\newcommand{\hlf}{\mbox{$\frac{1}{2}$}}

\newcommand{\ri}{\mbox{${\rm i}$}}
\newcommand{\rri}{\text{i}}

\newcommand{\beq}{\begin{equation}}
\newcommand{\eeq}{\end{equation}}
\newcommand{\hfp}[1]{\mbox{$\frac{#1}{2}^+$}}
\newcommand{\hfm}[1]{\mbox{$\frac{#1}{2}^-$}}

\newcommand{\aS}{\mbox{$\arg{S_l}$}\,}
\newcommand{\mS}{\mbox{$|S_l|$}}
\newcommand{\etal}{\emph{et al}}

\def\nuc#1#2{\relax\ifmmode{}^{#1}{\protect\text{#2}}\else${}^{#1}$#2\fi}
\begin{document} \graphicspath{{figures/}}

%\def\nuc#1#2{\relax\ifmmode{}^{#1}{\protect\text{#2}}\else${}^{#1}$#2\fi}
% To use, either will work $\nuc{16}{O}$ \nuc{16}{O}
%
%\begin{document} \graphicspath{{figures/}}

\title{The angular momentum dependence of nuclear optical potentials} \author{R.
S. Mackintosh} \email{raymond.mackintosh@open.ac.uk} \affiliation{School of
Physical Sciences, The Open University, Milton Keynes, MK7 6AA, UK}

\date{\today}

\begin{abstract}

The nuclear optical model potential (OMP) is generally assumed to be independent of the orbital angular momentum, $l$, of the interacting nuclei.  Nucleon-nucleus and nucleus-nucleus interactions are customarily  $l$ independent in calculations of nuclear elastic scattering and in standard reaction codes.  The evidence for various forms of $l$ dependence of OMPs is reviewed and the importance of implementing these forms is evaluated. Existing arguments and evidence for $l$ dependence are reviewed and new arguments and calculations are introduced. The relationship is examined between (i)
$l$ dependence, and, (ii) the undularity  (waviness) of  $l$-independent potentials  that are $S$-matrix equivalent to $l$-dependent potentials. Such undularity is a property of the dynamic polarisation potential (DPP) generated by the coupling to reaction channels, or by coupling to excited states of the target or projectile nuclei.  Various examples, particularly involving weakly bound projectile nuclei, are presented. Such undularity also occurs in $l$-independent potentials that have been found in model independent fits to precise wide angular range elastic scattering angular distributions.  Cases of such phenomenological undularity, for both light and heavy ions, are referenced and shown to be related to undulatory properties of the dynamic polarisation potentials (DPPs) arising from channel coupling.   Other forms of $l$ dependence, that could be standard options in direct reaction codes, are noted. Of particular importance are $l$ dependencies arising from full antisymmetrization. The case is made that reaction-induced $l$ dependence is a general property of nucleon-nucleus and nucleus-nucleus interactions and represents a valid extension of the nuclear optical model.

%Version 8e, in progress

\end{abstract}


\maketitle

\tableofcontents

%\tableofcontents
%
\newpage \section{INTRODUCTION}\label{intro}
The phenomenological optical model potential (OMP) for nucleon-nucleus and nucleus-nucleus
scattering  is almost always taken to be independent of the partial wave orbital angular
momentum,  $l$;  for  an early exception see Ref.~\cite{darri}. However there are both
theoretical and phenomenological  reasons to believe that some degree of $l$ dependence
is a  general property of nuclear optical potentials. These reasons will be presented
in what follows.

There are cases,  mostly involving light nuclei, see Section~\ref{HI-antisymm}, where fully
antisymmetrized calculations reveal strong and explicit
$l$ dependence, and these will be discussed. However the major concern of what follows
is $l$ dependence for which the theoretical arguments are less direct.
There is also a difficulty  identifying empirical evidence: for any $l$-dependent potential
there always exists an $l$-independent  potential with the same complex S-matrix $S_l$  (or $S_{lj}$).
The $l$-independent representation of an $l$-dependent potential,
as specified below, will never have a smooth Woods-Saxon-like form, and is generally
undulatory. This is a complication for elastic scattering
phenomenology and  for attempts to develop a unified single-particle
nucleon-nucleus interaction for both bound and scattering energies.

There are strong grounds for the $l$ dependence of OMPs
for composite projectile nuclei,  from deuterons to \nuc{16}{O}, making angular momentum
dependence  a generic property of nucleon-nucleus and nucleus-nucleus interactions.
Apart from its  intrinsic interest, this must influence the analysis of direct reactions.
Little is known concerning the effect of  the $l$ dependence of OMPs when they are applied in direct
reactions, in contrast to the way that non-locality due to exchange is routinely compensated for.

The $l$-independent representation of an $l$-dependent potential referred to
above is just the  $l$-independent potential that has the same S-matrix $S_l$ or  $S_{lj}$.  For nucleons
or other spin-\hlf\  projectiles, this potential is found from the $S_{lj}$
of the $l$-dependent potential by applying $S_{lj} \rightarrow V(r) + {\bf l \cdot s}\, V_{\rm SO}(r) $
inversion,  where $V(r)$ and $V_{\rm SO}(r) $ are $l$-independent. Inverting $S_l$ for spin-zero projectiles is straightforward, and some cases of inversion for spin-1 have been  carried out. Such S-matrix inversion will
be referred to explicitly or implicitly at many points in this work. Early studies of $S$-matrix inversion
are in Refs.~\cite{inv1,inv2,inv3};  for more recent reviews see
Refs.~\cite{s-pedia,Kuk04,CM89,arxiv}. The $l$-independent equivalent, found by inversion, will
always be characterized by a certain degree of undularity (waviness). That such undularity does
occur in precision fits to high quality scattering data, and that, furthermore, such undularity can be shown
to arise from channel coupling effects, will be a recurrent theme in what follows. The relationship between $l$ dependence and unuclarity is review idn Ref.~\cite{relation}. All the inversions
reported herein employed the iterative-perturbative, IP, inversion
method~\cite{inv1,inv2,inv3,s-pedia,Kuk04,CM89}.

For nucleons and other light ions,  there are global optical model  potentials that
describe elastic scattering data fairly well for a wide range of energies and target nuclei.
For nucleons see~\cite{KD}, for deuterons see~\cite{DCV} and for mass-3 nuclei see~\cite{Pang}.
Recently a theory-based global potential for nucleons of more than 40 MeV has appeared~\cite{furu}.
None of these global potentials include $l$ dependence. Such global potentials, which fit a lot of data fairly well, leave plenty of scope for $l$-dependence of potentials that precisely fit scattering data for each case. Such precise fits to data illuminate two aspects of elastic scattering: (i) the manner in which departures
of more exact fits from the global fit depend upon the nature of the target nucleus, and,
(ii) the occurrence and nature of $l$ dependence. As will become apparent, these two
are closely
connected.  The important point is that the difference between  a fit for which
$\chi^2/F \sim1$, and  a `\emph{good fit}' as commonly described, is a generally untapped source of information,
including evidence for $l$ dependence. One theme of this review is what might be learned from precise
fits to elastic scattering data; for a general discussion of this topic see Ref.~\cite{epja}.

Historically, the possibility  of a potential
model description  of nuclear scattering was for long considered  implausible. When it was found that
very simple  models~\cite{FPW,WS} could account  for much data in an approximate way, it became clear, see
e.g.~\cite{gww},  that  those simple models represented important understanding. However, although much
better fits to data have become possible, there has been little motivation to extract all the information
that might exist in precise, wide angular range,  elastic scattering data. In contrast, precise
charge densities have been extracted  from elastic electron scattering data, although
charge density is much less a model concept than the OMP. In fact,  almost all nuclear
elastic scattering data is incomplete, in more ways than one, as we shall make explicit with
examples. Where  precision OMP fits  to relatively complete data have been accomplished,
potentials with undulatory (wavy) features have emerged. These features lack a direct
explanation in terms of folding models and apparently depart from the original
concept of an OMP.

Undulations found in certain precision fits to elastic scattering data resemble those
arising as a result of coupling to reaction channels~\cite{relation}.  Such undulations occur in the
$l$-independent potential found by inversion of the elastic channel $S$-matrix from
coupled channel, coupled reaction channel or continuum discretized coupled
channel calculations. Such undulations also occur in the $l$-independent potentials
that are $S$-matrix equivalents of explicitly $l$-dependent nucleon-nucleus or nucleus-nucleus
that have been fitted to elastic scattering data~\cite{relation}.  An extension of the OMP to include
$l$ dependence thus makes it possible to exploit precision data in a way
that links it to the dynamics of elastic scattering.

{\bf Terminology used.} In what follows, CC indicates coupled channels in general, CRC (coupled reaction channel) as well as CDCC (continuum discretized coupled channel) for breakup channels. In the literature, `$l$-dependent' has sometimes been used to describe
potentials that are parity-dependent, parity dependence being a particular form of $l$ dependence
in which the potential takes the form $V_{\rm W}(r) + (-1)^l V_{\rm
M}(r)$ where the W and M subscripts label the (complex) Wigner and Majorana components.
In this work  `$l$ dependence' refers to any dependence of the OMP upon
the partial wave angular momentum $l$, including parity dependence.

We refer to potentials having the same $S$-matrix, $S_{lj}$ (and thereby the same observables),
as `$S$-matrix equivalent'. Every $l$-dependent potential has an $l$-independent $S$-matrix
equivalent that can be found by inversion.

{\bf Outline}. Section~\ref{theor} explains why, on the basis of standard theories, the OMP might
be expected to depend on angular momentum. Much of the discussion is based
on light ion, mostly nucleon, scattering. The nucleon OMP has
particular significance as a continuation into positive energies of the
shell model single-particle potential. Section~\ref{ccimply} surveys cases in which
the contribution to the OMP due to channel coupling has been determined, with results
that, it is argued, imply dynamically induced $l$ dependence for (mostly
nucleons and loosely bound nuclei. Section~\ref{implications} discusses
the extent to which $l$ dependence has been, and might be further, related
to experiment. Although the discussion up to this point
mostly relates to nucleon scattering, we argue that angular momentum
dependence is a general property of nucleus-nucleus interactions.
Section~\ref{composite} presents examples from the scattering
of  heavier nuclei connecting $l$ dependence to theory and experiment.
Section~\ref{imp-app} discusses what the existence of $l$ dependence
implies for phenomenological applications of OMPs. Section~\ref{conc} presents
general conclusions.

\section{ $l$ dependence and  theories of the OMP}\label{theor}
In the background to any discussion of $l$ dependence are various theories of the OMP.
Two well-developed theories are (i) due to
Feshbach~\cite{feshbach,feshbach1} and, (ii), going back to Bell and Squires~\cite{bellsq},
a theory based on the self-energy of a nucleon in nuclear matter. The latter has
especially been developed by Mahaux and collaborators~\cite{JLM,MS}, see
also~\cite{BR}. As emphasized by Mahaux and Satchler~\cite{temporal} there are
fundamental differences between these two approaches, not the least being
that there is no self-energy theory for composite particles.  Feshbach's
approach has long given insight into the scattering potentials for
composite systems but applications have generally been rather phenomenological
and, although exchange processes have been considered formally~\cite{feshbach1},
fully antisymmetrized applications are difficult to implement.

Other reaction theories such as the resonating group model, RGM,
also contribute to our understanding of interactions between lighter
composite nuclei, particularly with the exploitation of $S$-matrix-to-potential inversion.
RGM and related theories~\cite{WT,DD,SLYV}  include antisymmetrization exactly,
enabling them to reveal information concerning parity dependence, as well as
more general forms of $l$ dependence,  that arise from antisymmetrization.

\subsection{Feshbach theory}\label{fesh}
The theory of Feshbach~\cite{feshbach}
has occasionally been employed in calculations of the total contribution to the
nucleon OMP of all the channels that are coupled to the elastic channel, see
e.g.~\cite{rao,coulter,rawit87}. However, more commonly a form of this theory underlies
calculations of the contributions of specific selected channels to the OMP, for
example in various cases where it is apparent that particular processes are not
represented in conventional calculations~\cite{ghj,rsm71,satchler}.
It can represent processes that vary with nuclear properties in a
way that is excluded from the smoothly varying OMP of
standard folding models. Such varying contributions are identified as the
`dynamic polarization potential', DPP, see e.g.~\cite{satchler}.
In  the formal Feshbach theory the OMP or DPP are explicitly
$l$-dependent and non-local, as manifest in
Ref.~\cite{rawit87}. Nevertheless, local and
$l$-independent representations of the non-local and $l$-dependent DPP can be
found by S-matrix inversion; for recent examples see
Ref.~\cite{mkPRC81,pca40-2012,erratum,mk90}; Ref.~\cite{erratum} is an erratum for Ref.~\cite{pca40-2012}.
In general, local and $l$-independent
potentials representing DPPs exhibit undulatory, `wavy', features. Such features can
be compared with features of $l$-independent potentials that are $S$-matrix
equivalent to explicitly $l$-dependent potentials having known $l$ dependence.
Invariably,  local equivalent DPPs that represent coupling to specific
channels can \emph{not} be represented as a uniform factor multiplying the `bare' potential
without the coupling.  In many cases, DPPs can  be reliably determined far
into the overlap region of the interacting nuclei and, for nucleon scattering,
over the whole radial range.

Although  the many  contributions to the full
Feshbach OMP might somehow average to an effectively $l$-independent
potential, specific contributions to strongly coupled channels vary with the
target nucleus and depend upon the $l$ transfer, Q-values, etc. It is implausible that these
do \emph{not} lead to $l$ dependence and implausible that such $l$ dependence can
be represented as a local form varying smoothly over a range of nuclei
and energies.

\subsection{Potentials derived from self-energy}\label{self} Nucleon-nucleus
potentials due to Mahaux and collaborators~\cite{JLM,MS} and their later
extensions, Refs.~\cite{jlmb1,jlmb2}, `JLM potentials', provide a satisfactory,
but not precise, fit to nucleon elastic scattering data over a wide range of
energies and target nuclei. The formalism~\cite{JLM,MS} itself includes a local
equivalent to the specific non-locality that arises from knock-on exchange, the
major source of energy dependence of the JLM potentials. For a given energy, the
JLM  complex potential $V(r)$ depends on just the proton and neutron densities
at radius $r$: the local density approximation LDA. The original local density
model of Ref.~\cite{JLM,MS} was modified (the `extended local density
approximation' of Refs.~\cite{jlmb1,jlmb2}) in order to correct in a
phenomenological way the radial extension of the potential. When applied, this
model requires overall normalization factors which vary in a regular way.

Although the local density approximation was `extended'~\cite{JLM,MS} to correct
the radial size of the potential, it is still a local density model,
based only on the nucleon densities of the nucleus, and not, for
example, the density gradient. Specific properties of the nucleus such as
its collectivity do not enter and nothing in the model
corresponds to the orbital angular momentum of the interacting nucleon. Fits
to data do not in general, approach $\chi^2/F =1$, even with uniform normalization
factors which, as mentioned above, cannot accurately represent channel coupling. The model leaves
room for $l$-dependent terms.

\subsection{Nuclear Structure Approach}\label{NSA} The Nuclear Structure
Approach of Vinh Mau~\cite{VM}  and others~\cite{vmb,owm,wmc,mo} incorporates some of the physics of
the Feshbach approach with the self-energy method. In particular, it includes
the effect of coupling to particle-hole states corresponding to giant
resonances, and the relationship of this to $l$ dependence will be mentioned
later. The effect of such resonances has been incorporated into optical model
studies by Pignanelli \etal~\cite{pig81} and Delaroche \etal~\cite{del86}, and
see also~\cite{honorePRC33}. The calculations of Ref.~\cite{mo} are of particular interest
since they recognize the generation of $l$-dependence within the model, and we shall
refer to them again below. These calculations involve much the same physics
as the calculations of Rawitscher~\cite{rawit87} mentioned above.

We mention here that the optical model is discussed in the important monograph of
Mahaux and Weidenm\"uller~\cite{MW} who make explicit an approximation of their model
which appears to be implicit in calculations based on the nuclear structure approach.
This assumption is that there is just one nucleon in the continuum. This would exclude
a contribution from processes such as the coupling to deuteron channels. The contribution
of coupling to pickup channels can be,
and has been, studied within an extended Feshbach model~\cite{coulter}, as discussed below.

\subsection{Limits of local density models}\label{limits} Nothing in models
based on the local density approximation corresponds to $l$, the orbital angular
momentum of the nucleon. There is no scope for the $l$ of a scattering nucleon
to influence a nuclear interaction within models in which the finite size of the
nucleus and the density gradients in the nuclear surface enter only through the
way that the interaction at radius $r$ depends on the density around $r$.
The excitation of
inelastic channels involves projectiles in coupled channels propagating in
potential gradients around the nucleus. The coupling leads to non-locality
and $l$ dependence that
are not represented within the LDA. In Austern's picture of non-locality~\cite{austernpr137}, flux
leaves from the elastic channel at
one location and is restored at another location. This second location will in
general have a different local density. In the temporal non-locality discussed
by Mahaux and Satchler~\cite{temporal}, one can assume that the projectile will
return to the elastic channel after it has propagated to a region of different
density.  The non-locality and $l$ dependence that are
due to collective channel coupling will have effects on direct reactions, and
these effects have been studied in Ref.~\cite{km90}.

The density gradient in the nuclear surface plausibly leads to $l$ dependence.
Consider an incident nucleon interacting with a target nucleon in the surface,
where there is a nuclear density gradient. The reaction back on the incident
nucleon depends upon whether the target nucleon recoils into an
increasing or a decreasing nuclear density. This suggests a new term in
the potential proportional to $\vect{k} \vdot \nabla \rho{(\vect{r})} $, i.e.\
for a spherical nucleus,
\beq V_{\rm k}(r) =\vect{k} \vdot \vect{r} \pfd{\rho(r)}{r}. \label{grad} \eeq
Here
$\hbar\vect{k}$ is the local nucleon momentum. At high enough energies, where
the eikonal approximation is good, $\hbar\vect{k}$ may be taken as the incident
momentum, in which case a $\vect{k} \vdot \nabla \rho{(\vect{r})} $ will make a
zero contribution, as can be seen from the eikonal integral for
interaction $f(r) \vect{k} \vdot \vect{r}$:
\beq \chi(b) = - \frac{1}{\hbar v}
\int^{\infty}_{-\infty} f(r) \vect{k} \vdot \vect{r} {\rm d} z, \label{eik}\eeq
identifying $ k b$ with $l + \hlf$  and $S(b)= \exp{ ( \imath  \chi(b))}$ as usual.
However, the relationship to
$l$ dependence is immediately apparent with the help of Pythagoras:
\beq l^2 = k^2 r^2 - (\vect{k} \vdot \vect{r})^2 \label{lsq} \eeq
which holds when $\hbar
k$ is the local momentum. This implies that
\beq \vect{k} \vdot \vect{r} = \pm \sqrt{k^2 r^2 - l^2} \label{kdotr}\eeq
where the plus and minus signs apply at
the outgoing and incident sides of the target nucleus. These effects would not
cancel where the projectile is substantially absorbed or where the eikonal
approximation fails. If such a term were effective, it would constitute a source
of $l$ dependence. The formulation would be more elaborate if a
self-consistently calculated complex local momentum were to be included in a
term of the form $\vect{k} \vdot \nabla \rho{(\vect{r})} $.

\subsection{Implications of channel coupling}\label{cc} A long history of
calculations reveals that coupled channels, including reaction channels, make a
substantial contribution to elastic scattering. The contribution of low
lying vibrational states  to the proton OMP  was studied by Buck~\cite{bb63} and
Perey~\cite{fgp63}, and the contribution of rotational excitations of deformed
nuclei, to the OMP for other projectiles, was studied in Refs.~\cite{ghj,rsm71}.
For proton scattering, it was found~\cite{PL44B,NPA230} that coupling  to
deuteron channels by neutron pickup substantially modified the calculated
observables, in one case~\cite{kobmac76} greatly improving the fit for 30.3 MeV
protons on \nuc{40}{Ca}, a notoriously hard case to fit  (c.f.\ Section~\ref{Ca40});
the deep minimum in the angular distribution around 140$^{\rm o}$ was fitted. Later
calculations, in
which various approximations were lifted, reduced the effect, although the
recent studies~\cite{pca40-2012,erratum,remark,tobepub} of this case still reveals a substantial DPP
arising from the coupling to deuteron channels.  Although the deep minimum near
140$^{\rm o}$ is no longer fitted, the coupling has a large effect on the
angular distribution. The radial form of the DPP  is very
far from  a uniform renormalization of the `bare' (folding model)
potential: the real part is attractive close to the nuclear center, with a repulsive
region further out. The imaginary part is absorptive towards the nuclear
center becoming almost emissive in the nuclear surface; the real and imaginary
spin-orbit terms of the DPP are wavier than the central terms. Such
waviness is characteristic of $l$-independent representations of $l$ dependence.

As mentioned in Section~\ref{NSA}, Refs~\cite{pig81,del86} studied the effect on
proton elastic scattering of coupling to high-lying giant resonances.  This
coupling led to quite a good fit to the backward angle minimum for scattering from
\nuc{40}{Ca}~\cite{pig81} and also from \nuc{16}{O}~\cite{del86}; such coupling
should be studied together with pickup coupling. The effect of the
giant resonances must  be present for all target nuclei, not just closed
shell nuclei \nuc{16}{O} and \nuc{40}{Ca}. For these nuclei
the effect becomes apparent because there exist  deep minima
in the elastic scattering angular distributions which are not
filled in by the many active processes. The systematic contribution of giant resonance
coupling to $l$ dependence is not yet known. This contribution to
the OMP is likely to vary with energy and target nucleus in a different way
to the contributions of  low-lying collective states and particle transfer.
A recent study of the DPP
for protons coupled to both low lying and high lying collective states, is
Ref.~\cite{mk90}; the equivalent local  potentials are very undulatory,
indicating $l$ dependence.

A more complete listing of DPP calculations leading to undulatory DPPs for nucleon
scattering is given in Section~\ref{survey}.

\subsection{Relating coupled channel effects to $l$ dependence}
Ref.~\cite{mackob79} compared the effects on the elastic scattering
$S$-matrix,  $S_{lj}$, that are due to (i),
coupled neutron pickup  channels, Ref.~\cite{kobmac76},  with (ii), the
contribution to $S_{lj}$  of a phenomenological $l$-dependent term. This
comparison indicated that part at least
of $l$ dependence can be attributed to coupling to pickup channels. Subsequently,
coupled reaction
channel (CRC) calculations are much more highly developed enabling much
more rigorous comparisons of the same kind involving
explicitly $l$-dependent potentials fitted to data.
It is now straightforward to invert $S_{lj}$, for a given $l$-dependent
potential, as well as  from any CC calculation,  making it
possible to compare the resulting  $l$-independent potentials and thus match
empirical $l$ dependencies with $l$ dependencies arising from
channel coupling.

Delaroche \etal~\cite{del86} examined the effect of coupling to giant resonances
upon $|S_{lj}|$ but not upon $\arg (S_{lj})$. As shown in~\cite{kobmac76}, it is
the argument of the $S$-matrix which relates most directly to the effect on the
real part of the potential, especially for nucleons. The combination of $l$ transfer and momentum
transfer involved in exciting giant resonance states is a likely source of
$l$ dependence, and this deserves exploration.

\subsection{The contribution of knock-on exchange}\label{knock-on} It is
generally believed that knock-on exchange, represented by the Fock term in the
interaction between a scattered nucleon and the bound nucleons,  is responsible
for most of the energy dependence in the effective local nucleon-nucleus
interaction, the nucleon OMP. The contribution of knock-on exchange is included as a
local approximation in standard folding models, e.g.
Refs~\cite{JLM,MS,BR,jlmb1,jlmb2}. The explicit inclusion of knock-on exchange
requires the solution of integro-differential equations, e.g.
Ref.~\cite{OS,georgiev,kim1,kim2}, and is seldom carried out. The phenomenological
non-local potential of Perey and Buck (PB), Ref.~\cite{PB}, accounts for the
energy dependence of the local nucleon OMP, as shown explicitly in
Ref.~\cite{RSMSGC} in which $S_{lj}$ for the PB non-local potential was inverted
to produce the local equivalent. The inverted potential in this case did not
appear to be $l$-dependent suggesting that knock-on exchange does
not lead the $l$ dependence. However, Lukaszek and Rawitscher, Ref.~\cite{LR},
have shown that a realistic non-local exchange term, derived from a
non-local nuclear density, gives rise to $l$ dependence in the
nucleon-nucleus  local interaction. This $l$ dependence does not appear to be
parity dependence and deserves further study. The calculations of
Ref.~\cite{LR} apparently reveal a limitation of the symmetrical form of
non-locality of  Perey and Buck.  Nuclear matter calculations show that,
in a \emph{uniform} medium, exchange non-locality does indeed~\cite{JandM}  have
the Gaussian  form of Perey and Buck; however it is just the existence of a nuclear surface
that opens the possibility of $l$ dependence, see also  Ref.~\cite{mo} mentioned
in Section~\ref{NSA}.

\subsection{The special case of parity-dependence}\label{parity}
Particular exchange processes, especially in scattering
between light nuclei, give rise to parity dependence as a consequence of certain
exchange terms, including heavy-particle stripping.
The exchange terms arise from the antisymmetrization of projectile
and target nucleons, explicitly represented in resonating group model, RGM, calculations.

Parity dependence has been studied by fitting data, see Section~\ref{parityfits} and
Section~\ref{SGC}. Strong parity dependence was revealed in some cases.
Support comes by inverting $S_{lj}$
from RGM calculations for proton scattering from nuclei from mass 4 to
mass 40 as reviewed in \cite{s-pedia} and discussed below.
Baye~\cite{BayeNPA460} has presented theoretical arguments for the way in which
the strength of parity dependence is related to the masses of two interacting
nuclei. If one of these is a nucleon, then the Majorana terms become smaller as the mass of the
target nucleus increases. These predictions are borne out by studies of two
complementary kinds (see Ref.~\cite{s-pedia}): (i) $S$-matrix to potential
inversion of $S_{lj}$ found by R-matrix and other fits to scattering data,
and, (ii) S-matrix to potential inversion of $S_{lj}$ from RGM calculations. For
nucleon scattering from $^4$He, the same general result follows from studies
of type (i) and type (ii): that is, the odd-parity real potential has both volume
integral and RMS radius substantially greater than the even-parity potential. Ref.~\cite{s-pedia}
describes many other cases of nucleus-nucleus scattering.
As Baye predicted, the strength of the Majorana term for proton
scattering  falls off with the mass of the target nucleus, but is still
substantial for nucleon scattering from $^{16}$O, as found also in
Ref.~\cite{vos79} although that work was based on
imperfect fits to data. We return to parity dependence of nucleons on \nuc{16}{O} in
Section~\ref{SGC} where strong  evidence for it together with evidence for
coupling-induced $l$ dependence is  discussed. None of the parity dependent
potentials from studies of type (i) or type (ii) have an overall factor
$(1 + \alpha (-1)^l)$,  the standard assumption of purely phenomenological
fits,  Section~\ref{parityfits}.

Parity dependence leads to the enhancement of  differential
cross sections at backward angles. This  is often due to
heavy particle stripping in the case of nucleon scattering or cluster transfer
in interactions between heavier nuclei. One example of the latter is alpha
particle transfer in the case of \nuc{16}{O} scattering from \nuc{20}{Ne}. For
alpha particle scattering from \nuc{20}{Ne}, Michel and Reidemeister~\cite{MR}
showed that a small Majorana term markedly improved the fit to elastic
scattering angular distributions, apparently due to knock-on exchange of an
$\alpha$ cluster. A problem occurs in establishing
parity dependence: $S_l$ that originates from the parity-dependent
potential can always be fitted, by $S_l \rightarrow V(r)$
inversion~\cite{s-pedia}, with an $l$-independent potential. In the $\alpha$ plus
\nuc{20}{Ne} case the $l$-independent potential was found~\cite{cmzp}, in spite of the
smallness of the Majorana term, to be markedly undulatory and would not have
been discovered by standard fitting of the elastic scattering angular distributions.

In nucleon scattering from \nuc{6}{He}, the parity
dependence is, as expected~\cite{BayeNPA460}, less than for nucleon scattering
from \nuc{4}{He}, but still substantial, see Ref.~\cite{npa742}. An extreme case
of parity dependence, as determined from RGM S-matrix elements, is \nuc{3}{He}
on  \nuc{4}{He}, see Ref.~\cite{NPA589}. This reference also presents
parity-dependent potentials for \nuc{3}{H} on \nuc{4}{He} from RGM-derived
S-matrix elements. Ref.~\cite{NPA592} presents RGM-derived potentials for
nucleons scattering from light target nuclei having non-zero spin. In such
cases the character of the parity dependence strongly depends
on the channel spin. Refs.~\cite{s-pedia,Kuk04} review further cases where
parity dependence has been well established from RGM $S$-matrices which
include exchange effects explicitly.
Although the origin of parity dependence makes it somewhat distinct from
other forms of $l$ dependence, the problem of understanding its $l$-independent
representation is part of the same formal problem as the
$l$-independent representation of dynamical $l$ dependence. Ref.~\cite{mkPRC81}
presents the $l$-independent equivalent for an $l$-dependent potential,
the real part of which had a factor $(1.0 + 0.05(-1)^l) $,  for 15.66 MeV protons incident
on \nuc{8}{He}. The imaginary part as well as the real part of the equivalent potential had
undulations, including a small excursion into emissivity in the imaginary part.

Model calculations~\cite{inv2}, in which $S_l$ for explicitly parity-dependent potentials were inverted, led
to oscillatory $l$-independent potentials. The volume integrals of the oscillatory potential were  virtually
the same as for  those including a Majorana  factor.  It was stated:``The fact that a large
Majorana term makes little difference to the volume integral of the equivalent $l$-independent
potentials undermines the logic of the common argument to the effect that the agreement between
$J_{\rm R}$ and $J_{\rm I}$ for global optical optical potentials and nuclear matter theory indicates
that the essential features of the optical model are well understood.'' See also Ref.~\cite{berlin}.

\subsection{$l$ dependence and causality}\label{causality}
Causality imposes a relationship between the energy dependencies of the real and imaginary components of the
OMP, see e.g.~\cite{NMS}. This must imply a relationship between the $l$ dependence of the real component
and the $l$ dependence of the imaginary component.  This implies that for each value of the conserved
quantity $l$ there is a dispersion relation linking the two components~\cite{ATMR}. The consequences of
this have not been consistently pursued, but they have been studied in one case as reported in Section~\ref{reduced}. It is plausible that $l$ dependence in either the real or imaginary term will
be  accompanied by $l$ dependence in the other term.

\section{Inferring $l$ dependence from channel coupling DPPs}\label{ccimply}
The way that particular coupled channels contribute to the phenomenological OMP is  of interest
since it relates relating OMPs to the specific properties of the target nucleus or the projectile.
The nucleus-to-nucleus variation of collectivity and transfer reaction strength must affect the OMP
in ways that are absent from standard folding models. The  DPPs that are deduced
for specific channel couplings allow dynamical non-locality and the effects of $l$ dependence
to be identified with particular reaction processes. For general
references concerning the  (DPP) generated by channel coupling
see Section~\ref{fesh}.

The general method of determining the DPP due to specific coupled channels, is as follows:
a CC calculation, including the specified channels, yields elastic channel $S$-matrix $S_l$ or $S_{lj}$.
Inversion of this  $S$-matrix yields an inverted potential $V_{\rm inv}$, which may be complex and
contain spin-orbit terms. Subtracting the diagonal potential of the CC calculation, $V_{\rm bare}$
(the `bare' potential) from  $V_{\rm inv}$ yields a local representation of the DPP, $V_{\rm DPP}$.
The inversion is usually very exact so that the sum $V_{\rm bare} + V_{\rm DPP}$ is a local potential
that exactly reproduces the elastic scattering $S$-matrix from the CC calculation. In the case that  the parameters
of the CC calculation are fitted to elastic scattering data,
$V_{\rm inv} = V_{\rm bare} + V_{\rm DPP}$  is a
local potential that fits that data with a local representation of the specific included channel
coupling effects. It is therefore appropriate for comparison with phenomenological local
potentials;  it contains a local and $l$-independent representation of the non-local and $l$-dependent
formal DPP. This formal DPP could in principle be calculated using Feshbach's formalism, Ref.~\cite{feshbach,feshbach1,rawit87,satchler}, but a conversion to a local and $l$-independent form, suitable
for comparison with phenomenological systematics,  would require considerable effort and yield a less exact form.

In Section~\ref{survey} we briefly describe fairly recent calculations of DPPs for scattering of nucleons
from a range
of nuclei from \nuc{6}{He} to \nuc{40}{Ca}. We omit discussion of earlier works that were carried out
before recent developments in codes and computers permitted more exact coupled reaction channel, CRC, calculations. There are some references to earlier works elsewhere in this paper as well as in the cited works.

Two general conclusions apply to all the results: (i) the DPPs are never proportional, point by point,
to the bare potential,
for an example see Fig.~\ref{fig1}.
Thus, fitting elastic scattering angular distributions by uniformly renormalizing folding model potentials is to be deprecated, and,  (ii) the DPPs are generally undulatory to varying degrees, often having radial
ranges where the imaginary term is emissive. In some cases even the full inverted potential, $V_{\rm bare} + V_{\rm DPP}$, has local emissive regions, but not breaching the unitarity limit $|S_L| \le 1$;  the CC calculation itself satisfies this limit. These properties will be relevant to the question of $l$-dependence.

In Section~\ref{loose} we review the DPPs arising from the interaction of loosely bound nuclei with target nuclei, and the above  two general properties apply in these cases. The inverted potentials and DPPs presented in Figs~\ref{deut-1} and~\ref{deut-2}, for the case of 52 MeV deuterons on \nuc{40}{Ca}, show coupling effects that are quite typical for energies well above the Coulomb barrier.

\subsection{Survey of DPP results for nucleon scattering}\label{survey}
\begin{figure} \caption{\label{fig1}  DPPs  generated by coupling to pickup channels for
30.3 MeV protons on $^{40}$Ca.  From top: real and imaginary central potentials,
real and imaginary spin-orbit potentials. The dashed lines are  for the bare potentials, solid lines for a full set of pickup states, and the dotted lines are for coupling to the lowest \hfp{3} state,
see Ref. ~\cite{pca40-2012,erratum,remark,tobepub}. }
\begin{center}
\psfig{figure=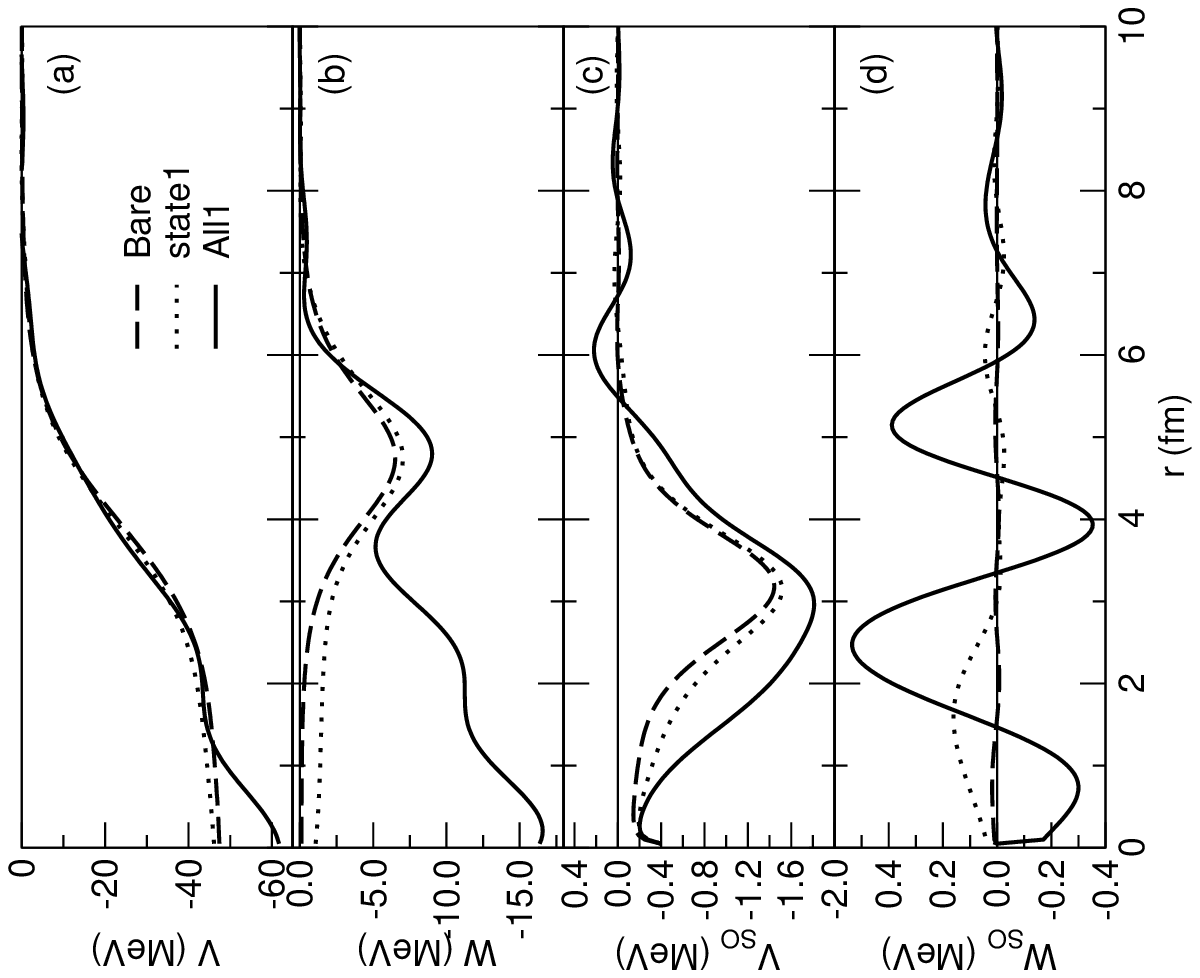,width=10cm,angle=-90,clip=} \end{center} \end{figure}

The contribution of neutron pickup channels to the proton OMP for scattering from \nuc{40}{Ca}
at 30.3 MeV was studied using CRC in Ref.~\cite{pca40-2012,erratum,remark,tobepub}.
The pickup coupling generated local potentials that are compared with the bare potential.
A conspicuous feature in the real central part is the deep attraction
induced by coupling  at the nuclear center. However, the overall effect is actually repulsive, with an
overall reduction in the volume integral of 7.4 MeV fm$^3$ i.e. \ $\Delta J_{\rm R} = -7.44$ MeV fm$^3$
(all quoted volume integrals are given as per nucleon
according to the convention of Ref.~\cite{satchler}.) The fact that the rms radius of this component was increased
by 0.034 fm is a result of undulations in the surface that are not apparent on the scale of  Fig.~\ref{fig1}.
The dashed lines present the bare potential, dotted lines are for coupling to just the lowest \hfp{3} state of
\nuc{39}{Ca}. The solid lines are for coupling to a large set of \hfp{1}, \hfp{3}, \hfp{5} and \hfm{7} states. Ref.~\cite{tobepub}
explores the development of the deep attraction at the nuclear center which is absent when fewer
states are coupled. The spin-orbit terms were conspicuously undulatory, as shown in the lower panels of  Fig.~\ref{fig1}.  For coupling to just the \hfp{3} state, dotted lines, the overall reduction in the volume integral of the real part was greater than for the solid line:
$\Delta J_{\rm R} = -12.88$ fm$^3$ and the rms radius was increased by 0.33 fm. Quite apart from the undulatory features, the change in rms radius shows that the coupling cannot be represented as a renormalization.

The imaginary  central DPP was also somewhat undulatory  in the surface, having a localized  region near 6.8 fm
where it is almost emissive. The real and imaginary spin-orbit terms are very undulatory.
In summary: the effects of coupling to pickup channels could certainly not be reproduced with
$l$-independent potentials having standard forms, or renormalized folding model potentials.
In particular, the undularity in the surface region implies $l$ dependence.

By contrast, the DPP arising from the coupling to vibrational collective states in the same nucleus at the same
energy, has strong undulatory features  in all components, Ref.~\cite{mk90}. With
a large set of vibrational states,  the amplitude of the undularities in the DPP are large enough
point-by-point to be disproportionate to the overall changes in the volume integrals.
Unlike the case of pickup coupling, the overall effect on the real central DPP, as measured by the volume integrals, is attractive.

The contribution to the interaction between protons and \nuc{6}{He} due to breakup of  \nuc{6}{He}, was studied in Ref.~\cite{prc67} at c.m.\  energies of 22 MeV and 35 MeV.  For a nucleus of such limited radial extent there is an obvious limit to the possible degree of undularity. The real central DPP is substantially repulsive  for $r$ less than $\sim 2$ fm and attractive in the surface. The imaginary central potential is absorptive
around 2 fm with a tendency to become emissive near the origin and in the surface. The real and imaginary spin-orbit DPPs both change signs from negative to positive at $r\ \sim 2$ fm. The form of $S_{lj}$ corresponding to this behavior could not occur with an $l$-independent potential without undulations.

The contribution to the interaction between protons and \nuc{6}{He} due to neutron pickup from  \nuc{6}{He} was studied in Ref.~\cite{KMprc83} at  71 MeV/nucleon. Pickup leading to the 1/2$^-$ and 3/2$^-$ states were included separately and together.  At this energy, the  DPPs  are more undulatory than for  the lower energy case~\cite{prc67} involving breakup of the \nuc{6}{He} target.  It was found that breakup of the outgoing deuterons significantly modifies the DPPs, reducing the repulsive effect on the real central term, but not enough to modify the overall finding: pickup coupling induces repulsion and absorption in the central term  and modifies the spin-orbit terms in a complicated way. The overall effect could not be simulated without $l$-dependence.

The contribution to the interactions between protons and \nuc{8}{He} due to coupling to neutron pickup channels  was studied in Ref.~\cite{skaza} and, more completely and over a wider energy range, in Refs.~\cite{mkPRC81,he8a}. The DPPs  for the most complete calculation~\cite{mkPRC81}, designated `d5' at 15.66 MeV/nucleon, have real central terms that were repulsive around 1.5 fm and 4.5 fm and attractive near 3 fm, i.e.\ it is undulatory. The imaginary central term is absorptive for all $r$, but peaked around 1.5 fm, i.e.\ well within the overlap region. The DPP  has substantial real and imaginary spin-orbit terms. At 25 MeV/nucleon, the general shapes of the various terms are similar, although the  repulsion  in the real central term for low $r$  is greatly enhanced near the origin. For the higher energy of 61.3 MeV/nucleon, the properties are generally similar again  but with a marked reduction in the wavelength of the undulations, which are quite substantial out to about 7 fm for the central terms. The spin-orbit DPPs are largely confined to $r\le 2$ fm. The relevant point once more is that the DPP could not possibly be well represented by a smooth and $l$-independent potential.  In addition, Ref.~\cite{mkPRC81} presented arguments that the pickup-induced DPP was dynamically non-local, consistent with theory, e.g.~\cite{feshbach,satchler}.

In Ref.~\cite{be10}, the contribution of neutron pickup to the proton-\nuc{10}{Be} interaction was studied at 5 energies between 12 MeV and 16 MeV. For the real central term, a consistent pattern of repulsion near 2.5 fm and attraction near 6 fm was found, with an overall strong repulsive effect as quantified by the volume integral. The appearance of a $0.5$ MeV deep attractive region at 6 fm, where the bare potential is very small, suggests that the coupling effect on the real central  component could  not be represented with a smooth $l$-independent form.  The imaginary central term showed a consistent pattern of emissiveness for $r < 2$ fm and absorption further out. At these lower energies, where undularities have a relatively long wavelength, a reasonable representation by a smooth potential may be possible.

In summary, for the cases described here, with the possible exception of the lowest energy cases (for protons on \nuc{10}{Be}), \emph{a potential model representation of the elastic scattering $S$-matrix requires either an undulatory potential or an explicitly $l$-dependent potential.} The  calculations described here do not identify the nature of the $l$-dependence. A complication is that the formal DPP is dynamically non-local as well as $l$-dependent.

\subsection{Scattering of loosely bound nuclei}\label{loose}
The first exact $S_l \rightarrow V(r)$ IP inversions~\cite{inv1}  were for breakup of \nuc{6}{Li}. The DPPs, exhibited strong repulsion in the surface region, showing explicitly why  the folding model for this nucleus required renormalization. Such  renormalization  was not required when the same folding model was applied to less loosely bound projectiles. An approximate inversion method, weighted TELP, had already been applied and came to the same general conclusion concerning the surface region. However, the weighted TELP method, although
approximately correct in the surface region,  misses various features, in particular undularities, that appear in exact inversion and which are relevant  to the question of $l$-dependence. Comparisons of  weighted TELP and exact inversion can be found in Ref.~\cite{pmPRC84}. Exact inversion can provide information concerning the DPP deep into the overlap region. This is significant as found ~\cite{GM-D} in connection with transfer DPPs for \nuc{8}{He}.

Ref.~\cite{universal} determined the DPPs due to the breakup of 80 MeV deuterons on \nuc{58}{Ni} and 156 MeV \nuc{6}{Li} on \nuc{12}{C}. The DPPs were well-determined down to small radii and showed remarkably similar distinctive radial shapes for deuterons and \nuc{6}{Li}. The similarity applied in the comparison of two breakup models: (i) when there was only S-wave breakup, and (ii) when D-wave breakup was also included. The strong undulatory shape of these DPPs was such that the real part varied with $r$ from substantial surface repulsion to  double peaked attraction within the overlap region. The elastic scattering $S$-matrix responsible for such shapes could not be reproduced with an $l$-independent potential. A later study~\cite{mpPRC86} of the breakup DPP for deuterons on \nuc{58}{Ni}, involving more precise breakup calculations for 56, 79 and 120 MeV deuterons, found essentially the same pattern
of attraction and repulsion in the real part, and of absorption in the imaginary part. One respect in which
the later calculations differed was in the more distinct region of emissiveness in the imaginary part. There is no possibility of the CDCC elastic channel $S$-matrix being reproduced by a smooth $l$-independent potential.

Ref.~\cite{pmPRC84} presented the DPPs due to the breakup of \nuc{6}{Li} on \nuc{12}{C} at 90, 123.5, 168.6, 210 and 318 MeV with parameters adjusted to fit elastic scattering angular distributions at each energy. At the highest energy, a notch test indicated sensitivity down to a radius of about 2 fm. The DPPs were consistent with those found in the earlier less rigorous calculations of Ref.~\cite{universal}. Ref.~\cite{pmPRC84}  also reported a comparison of weighted TELP inversions and precise $S$-matrix inversions, revealing the limitations of the former. In particular, weighted TELP failed to get the magnitude of the DPP correct in the outer radial regions and entirely missed significant structure at smaller radii, where the notch test indicated sensitivity.

There is much interest in the interaction of loosely bound nuclei with heavier nuclei at energies near the Coulomb barrier. This situation  presents difficulties for inversion calculations, but the study~\cite{kmb} of \nuc{8}{B}, \nuc{7}{Be} and \nuc{6}{Li}
on \nuc{58}{Ni} at energies near the Coulomb barrier did present DPPs for  \nuc{8}{B} and \nuc{7}{Be}. The DPP
for  \nuc{8}{B}  differed from that for \nuc{7}{Be}, but both the real and imaginary parts had general features in common. In both cases the real part had a very long attractive tail that became repulsive further in. In each case there was a long absorptive tail which for \nuc{7}{Be} had
a wavy feature, but became emissive for $r < 9$ fm whereas the imaginary part for \nuc{8}{B} remained absorptive. The case of 170.3 MeV \nuc{8}{B} on  \nuc{208}{Pb}, involving the  coupling to breakup channels, was more amenable to inversion, Ref.~\cite{mpPRC88}.  In this case the real DPP had long range attraction, becoming repulsive for $r<15$ fm. The imaginary DPP had a long absorptive tail, becoming emissive for $r<11$ fm. Currently achievable experimental elastic scattering angular distributions would  probably be insensitive to these details. Nevertheless, any smooth potential reproducing the CDCC elastic channel $S$ matrix would necessarily be $l$-dependent.

For 52 MeV deuterons scattering from \nuc{40}{Ca}, the combined contribution to the deuteron-nucleus potential due to the coupling to \nuc{3}{H} and \nuc{3}{He} pickup channels and also deuteron breakup channels,  was studied in  Ref.~\cite{kmPRC77}.  In certain cases, see below, coupling to proton channels (stripping) was also included. The selected processes were included together in CRC calculations in which the bare OMP was adjusted to optimise the fit to the elastic scattering angular distribution. The $(d,t)$ angular distributions were also fitted. When stripping was included, vector and tensor analysing powers were calculated, enabling the contribution of the coupling to the spin orbit and $T_R$ tensor interactions to be determined, although experimental data to compare with these are lacking. It was shown that the complex, central contribution to the DPP was well established, and that is briefly summarized here.

\begin{figure} \caption{\label{deut-1}  For
52 MeV deuterons on $^{40}$Ca,  real and imaginary central potential,
real part above and imaginary part below. The solid lines are the bare potential and  the dashed and
dotted lines are for two alternative inversions, as described in the text, of the elastic scattering $S$-matrix
when full channel couplings, apart from stripping, are switched on. }
\begin{center}
\psfig{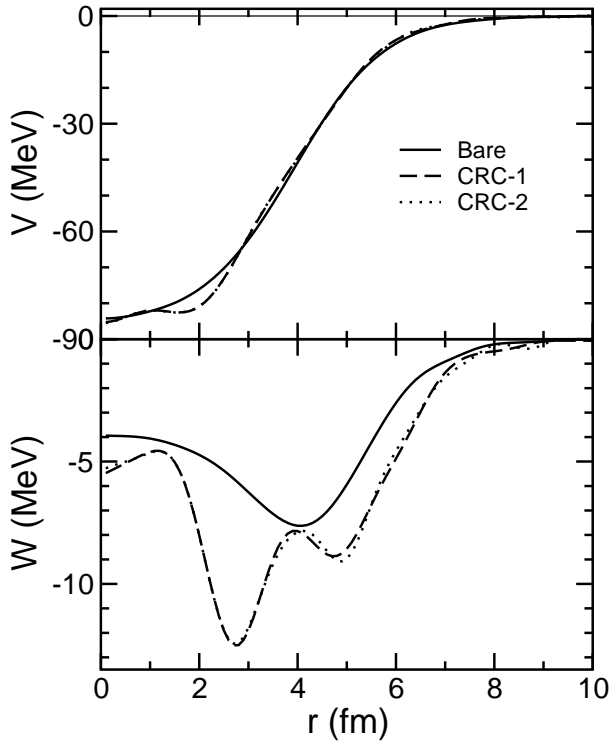} \end{center} \end{figure}

\begin{figure} \caption{\label{deut-2}  For
52 MeV deuterons on $^{40}$Ca,  real and imaginary central DPPs found by subtracting
the bare potential from the inverted potential,
real part above and imaginary part below. The solid and dashed lines correspond to the dashed and dotted lines
of Fig.~\ref{deut-1} and the dotted lines  are for a third, alternative, inversion, see Ref.~\cite{kmPRC77}.}
\begin{center}
\psfig{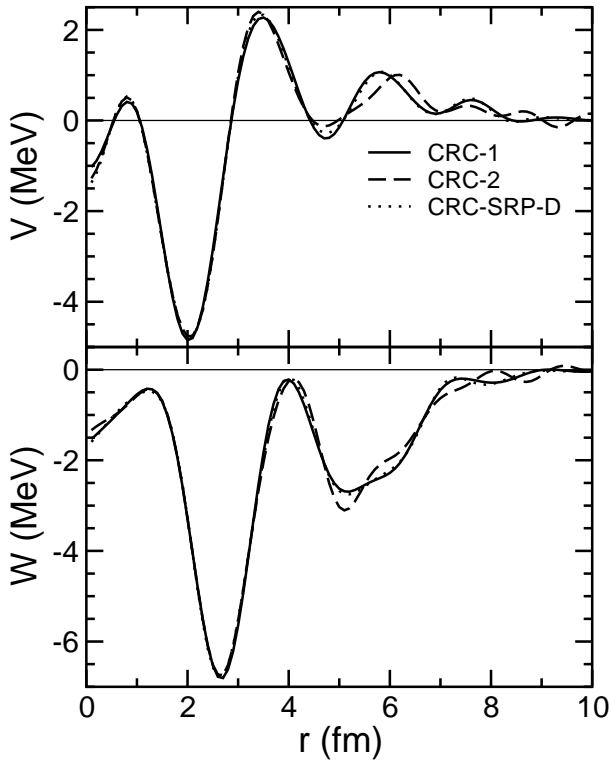} \end{center} \end{figure}

\begin{figure} \caption{\label{deut-3}  For
52 MeV deuterons on $^{40}$Ca,  real and imaginary central DPPs found by subtracting
the bare potential from the inverted potential,
real part above and imaginary part below. The solid line is as for the solid line in Fig.~\ref{deut-2}
and the dashed line is for the central terms of the `full', i.e.\  not $J$-weighted, inversion. }
\begin{center}
\psfig{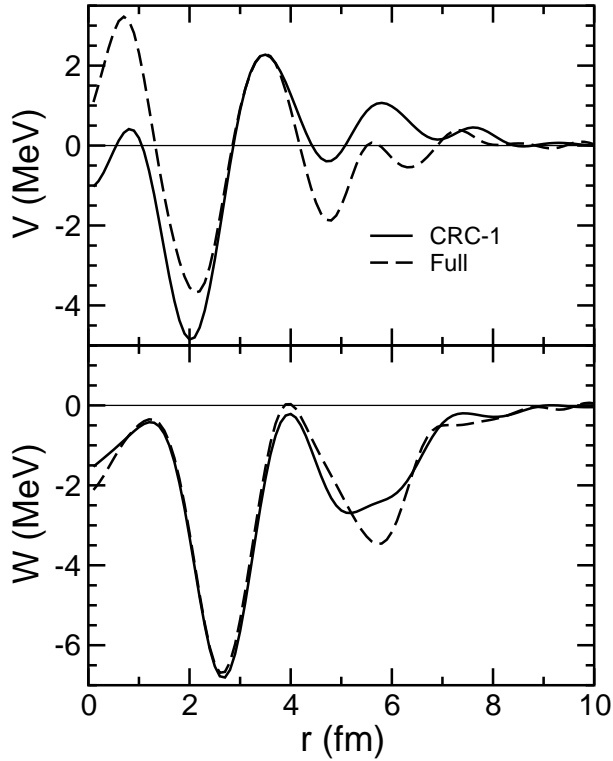} \end{center} \end{figure}

Although spin-1 inversion of  $S^J_{l'l}$ leading to the $T_R$ interaction is possible, and results of this were presented in Ref~\cite{kmPRC77}, it is also of interest to test the possibility of a suitable $J$-weighted inversion, which leads to reasonable central terms, and  is applicable in cases where full spin-1, or spin-$\frac{3}{2}$, inversion is unavailable. Fig.~\ref{deut-1} compares the bare potential with two inverted potentials calculated with $J$-weighted inversion. The inverted potential shown as dotted lines has a closer fit to the $S$-matrix (technically, a lower inversion-$\sigma$, see Ref.~\cite{inv2,s-pedia,Kuk04}). The contribution to the real part appears small in this figure, but when the bare potential is subtracted from the inverted potentials, it can be seen that the real DPP is comparable in magnitude to the imaginary DPP, as shown in Fig.~\ref{deut-2}.  In this case the imaginary DPP is quite substantial compared to the  imaginary bare term,
and is everywhere absorptive; this is by no means always the case with  reaction channel coupling.
The relevant point  is that the radial shape is strongly undulatory, as typical when reaction
channel coupling is included.  No smooth $l$-independent potential could reproduce the $S$-matrix.
The $J$-weighted inversion gives generally reasonable results as can be seen in Fig.~\ref{deut-3}
where the dashed lines represent the central potential from a full spin-1 inversion. The imaginary
part is little affected and actually becomes less absorptive at 4 fm, becoming slightly emissive there.
The real part is substantially modified for $r<2$ fm and also around 6 fm.

\begin{figure} \caption{\label{deut-4}  For
52 MeV deuterons on $^{40}$Ca,  DPPs for complete coupling including pickup.
From top, real central,  imaginary central, real tensor $T_R$, imaginary tensor $T_R$.  For the complex spin-orbit terms, see Ref.~\cite{kmPRC77}.
Included coupling: dotted lines, reorientation only; dashed lines add deuteron breakup; dot-dashed lines
add pickup to \nuc{3}{He} and \nuc{3}{H}; solid lines include also stripping to states of \nuc{41}{Ca}.  The dot-dashed lines for the central terms correspond to the dashed lines in Fig.~\ref{deut-3}.}
\begin{center}
\psfig{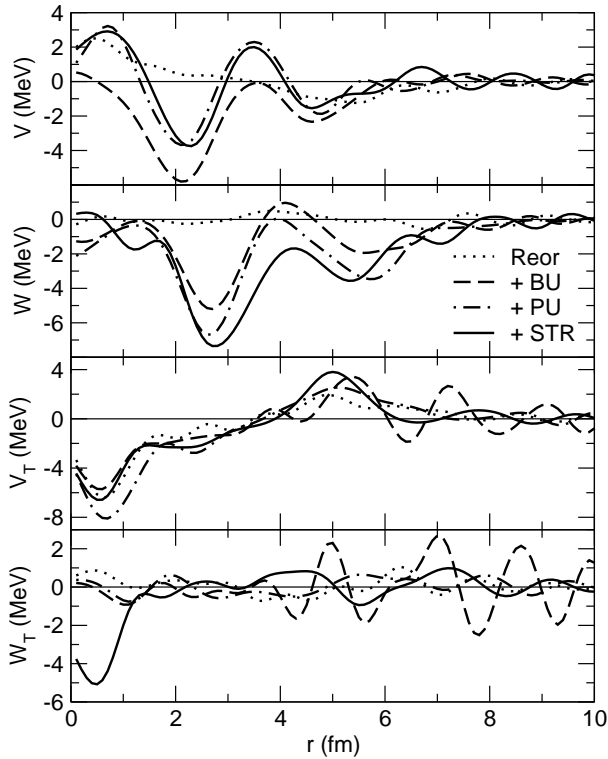} \end{center} \end{figure}

The DPP when stripping channels are included in a more complete CRC calculation is presented by the
solid lines in Fig.~\ref{deut-4}.  The  imaginary DPP is more absorptive for most values of $r$, but becomes
more undulatory in the surface region, with distinct excursions into emissiveness. The dot-dashed lines for the central terms in Fig.~\ref{deut-4} are the same as the  dashed lines in Fig.~\ref{deut-3}.

\subsection{Consequences for phenomenology}\label{consequences}
The generation of undulatory DPPs by channel coupling is a general phenomenon.  Empirical study
requires precise model independent fits to wide angular range elastic scattering
data.  Such model independent fitting should not be terminated at the point when precise fitting seems
to suggest undulatory potentials, c.f.\ Ref.~\cite{ARF}.   The evaluation of folding models must not simply
involve a search on a normalization constant to optimise the fit, but should determine an  additive
form by model independent fitting.

For the cases at energies near the Coulomb barrier, the properties of the DPP are rather abstract since the details will be very difficult to relate to experiment. Nevertheless, the properties found for such low
energy DPPs contribute to a systematic understanding of $l$ dependence generated by channel coupling.

\section{Experimental evidence for $l$ dependence}\label{implications}

\subsection{The problem of identifying $l$ dependence}\label{equiv}
There is a particular difficulty in
establishing $l$ dependence in a convincing way since any
S-matrix $S_{l}$, depending on partial wave angular momentum $l$, can be subject
to $S_l \rightarrow V(r)$ inversion~\cite{s-pedia,Kuk04,CM89,arxiv} (or $S_{lj}
\rightarrow V(r) + \vect{l} \vdot \vect{s}\, V_{\rm SO} (r)$ inversion; the
possibility of spin-orbit inversion is implicit when not stated) leading to an
$l$-independent potential. For example $S_l$ for an explicitly parity-dependent
potential can be inverted to yield a parity-independent potential. Any $l$-independent
potential  that represents an $l$-dependent potential in this way
will be undulatory, ranging from  mildly wavy to markedly oscillatory. Even if the two potentials
($l$-dependent and $l$-independent) yield the same $S_l$, they will, in general, have
different wave functions within the range of the potential.  In many contexts
an $l$-dependent representation is clearly more physical than a very undulatory potential.
Any elastic scattering data can be fitted by a local $l$-independent potential, perhaps
determined by model-independent fitting (sums of spline functions, Gaussian functions,
Bessel functions etc.) or by fitting $S_l$ to the data followed by inversion of $S_l$. In either case,
finding an undulatory potential probably indicates an underlying $l$-dependent
potential; examples will follow.

Although it is easy to find an $l$-independent  equivalent to any given $l$-dependent potential,
the inverse to this, i.e., identifying the specific form of $l$ dependence from empirical waviness,
is an open problem.  It is always possible to find complex coefficients $f(l)$ for any $V(r)$
such that,  for all $l$, $f(l) V(r)$ reproduces given $S_l$, but this will not be easily interpreted.

The wide range of possible forms of $l$ dependence makes establishing unique  $l$
dependence by fitting experimental data with $l$-dependent forms a daunting task. One approach is to
exploit the alternative representations of the potential (wavy or explicitly $l$-dependent).
In principle, model independent fitting should achieve a perfect ($\chi^2/{\rm DF} \sim 1$) fit
to observables that have been measured with high precision over a wide angular range. Such
precise  fits should provide uncertainties  and may yield wavy potentials (for 52 MeV deuterons,
see Ref.~\cite{ermer}, for 20 MeV deuterons see Ref.~\cite{ermer2}, for protons, see
Refs.~\cite{kmAnnP,ARF}) and `all' that remains is to establish a correspondence
between particular forms of undularity and corresponding forms of $l$ dependence.
Establishing $l$ dependence, as a signature of the limitations of the local density approach,
is of sufficient interest that the extraction of the full information content of elastic scattering data,
by precisely fitting data,  is a worthwhile objective. In fact, attempts to extract  the full information
content of elastic scattering data have become rare.  Claims for the `limitation of the one-channel
phenomenological optical  model'~\cite{shamu}, based on the failure to achieve fits with
(visual estimate) $\chi^2/N\sim 20$, are invalid.  A failure of Woods-Saxon, WS, potentials to achieve
$\chi^2/N\sim 20$, or even $\chi^2/N\sim 1$, is \emph{not} a failure of the phenomenological optical
model, but the failure of an unnecessarily restricted form of potential. It is always possible
to find a complex potential giving a perfect fit; the problem is one of interpretation.

The belief that it is worthwhile to extract the full information content from
hard-won, high precision elastic scattering data, appears to be less universal
than the belief, commonly implied in the literature, that a fit with
$\chi^2/N\sim 20 $ is `good'. What constitutes a `good fit' can be a matter of
context, but contexts certainly exist where it is appropriate to extract the
full information content of elastic scattering data. The question is just how to
extract all the information;  there will always be an $l$-independent equivalent
to any $l$-dependent potential that gives a precise fit to the data,
although certainly not of Woods-Saxon form. In fact, there may be
many potentials giving perfect fits, when, for proton scattering,
there is no measurement of the Wolfenstein spin-rotation R-parameter, see
Ref.~\cite{kmr}. Little is known of the topology of the
region in parameter space defined by $\chi^2/{\rm DF} \sim1$, for
data of specific quality, see Ref.~\cite{rsm79}. Thus, such properties of the
nucleon-nucleus interaction as its possible $l$ dependence can be hard
to establish  unambiguously, even from precise fits for a single nucleus at a single
energy: such fits are necessary but not sufficient. In this connection note that model
independent fits to proton scattering, presented in Ref.~\cite{ARF}, did not achieve  precise fit
to 30.3 MeV proton scattering from \nuc{40}{Ca}, missing exactly the feature at backward angles
that motivated the original $l$-dependent fits. This was evidently because very close
fits were not sought, presumably because of the resulting development of undulatory
characteristics. Another
model independent fit for this case, Ref.~\cite{kmAnnP}, applying `theoretically unprejudiced
fits' actually was at fault in that the following prejudice was applied: that the imaginary potential
must be nowhere emissive. It is now clear that this is not required for maintaining the unitarity limit,
and it is commonplace for DPPs generated by channel coupling to have emissive regions
without breaking the unitarity limit. Explicitly $l$-dependent potentials commonly have
$l$-independent  equivalents with emissive regions.

Finally, we note that most elastic scattering data is \emph{incomplete}.  Very commonly, there are no measured
values for significant parts of the angular range. This is often an unavoidable consequence of experimental
circumstances, but the consequences of the incompleteness should be recognized.
The restriction on angular range bears directly on the question of $l$ dependence for the case of
\nuc{3}{He} scattering, as described in Section~\ref{he3}. Moreover, as mentioned above, there are
almost no spin-rotation measurements for nucleon elastic scattering.

\subsection{Evidence from singular behavior at the origin}\label{cusp}
The undularity of $l$-independent potentials representing  $l$ dependence was established in
Ref.~\cite{inv2} in which inversion of $S_l$ for potentials with added  $l$-dependent terms led to
quite strong undulations. The waviness in the surface included
regions in the surface where the overall attractive potential $V(r)$ was actually repulsive, but,
significantly,  $V(r)$ had a strong radial derivative at $r=0$. As a consequence, the same potential  in three
dimensions, $V(\vect{r})$,  would have a marked cusp at the nuclear center. Where a potential $V(r)$
is reliably established by  phenomenology to have a non-zero derivative at $r=0$, then the non-physical
nature of a central cusp implies  that the potential must represent an underlying $l$-dependence. It is likely
that mostly just nucleon and deuteron elastic scattering analyses will have sufficient sensitivity to the
$V(r)$ near $r=0$  to establish a non-zero radial derivative in $V(r)$ there. The natural interpretation of
such a property  is that the potential is an $l$-independent representation of an $l$-dependent potential.

\subsection{Direct evidence for $l$ dependence from fits to nucleon elastic scattering}\label{fits}
\begin{figure} \caption{\label{cs1}  For
30.3 MeV protons on $^{40}$Ca,  the solid lines are the angular distribution
(above) and analyzing power for the $l$-dependent potential of
Ref.~\cite{cordero}. The dashed lines are calculated with the same potentials
except that the $l$-dependent component is omitted; the difference represents
the effect of the $l$-dependent component.}
\begin{center}
\psfig{figure=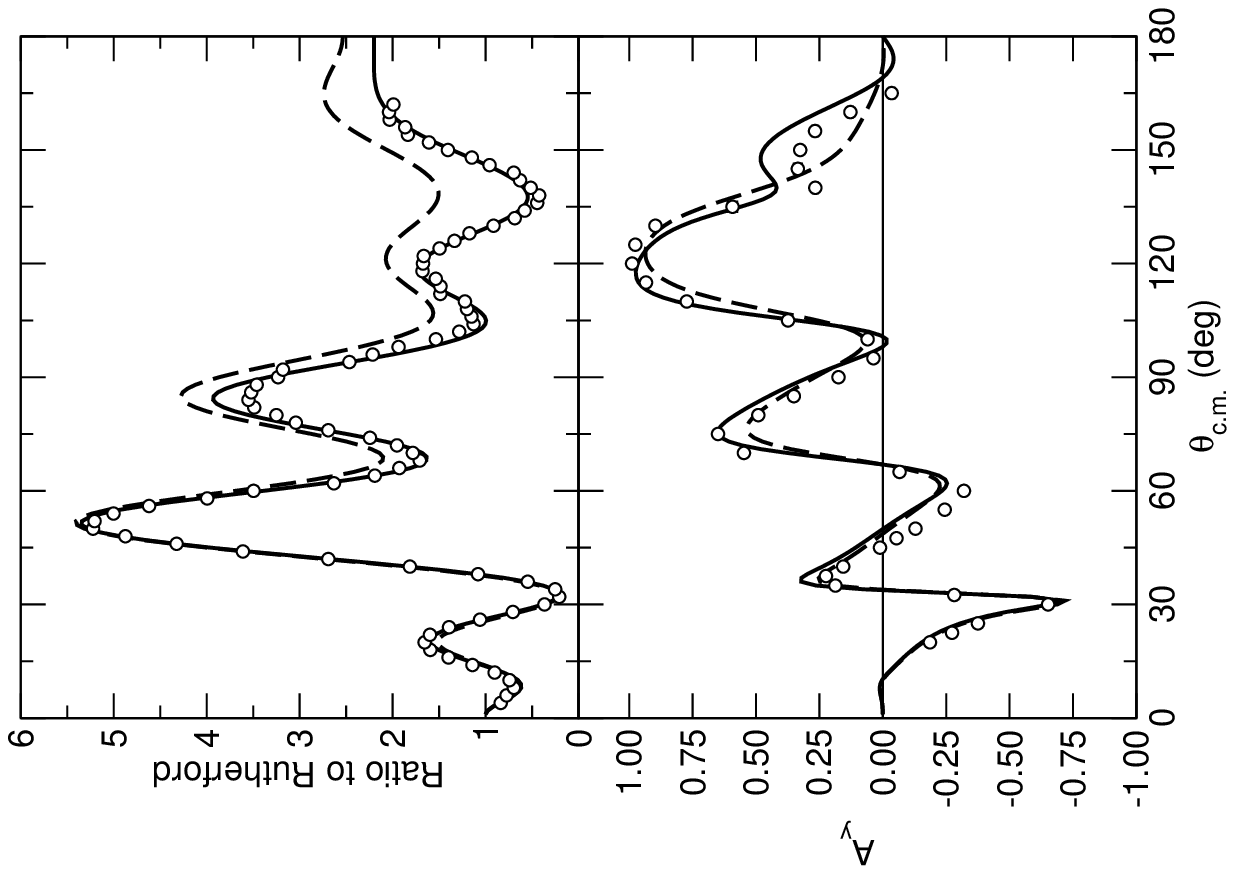,width=8cm,angle=-90,clip=} \end{center} \end{figure}

In Ref.~\cite{cordero} an $l$-dependent term was added to an OMP of standard
form leading to a substantial improvement to fit to the angular distribution and
analyzing power data for 30.3 MeV protons scattering from $^{40}$Ca. The data
were of unusual precision and of wide angular range and had resisted all
attempts to achieve $\chi^2/N$ less than about 10, see Section~\ref{interp}.
The $l$-dependent term, which was added to a standard 7-parameter WS plus WS-derivative
$l$-independent central potential, had the following $l^2$-dependent form:
\beq U_l(r) = f(l^2, L^2, \Delta^2) (V_l g_{\rm R}(r) + \ri W_l g_{\rm I} (r)) \label{lterm}\eeq
where the functions $g_{\rm R}(r)$ and $g_{\rm I}(r)$ are standard WS derivative
terms and $f(l^2, L^2, \Delta^2)$ is the standard WS form with $L^2$ being the `radius' and $\Delta^2$
the `diffusivity'. The spin-orbit component had no $l$-dependent term. The
$l$-dependent potential did fit the deep minimum  in the angular distribution
around 140$^{\rm o}$ that no WS (or folding model) potential has fitted.
Ref.~\cite{cordero} compared fits to the data by the $l$-dependent potentials
and the best WS $l$-independent potential. In Fig~\ref{cs1} we compare the
angular distribution and analyzing power as calculated from the $l$-independent
part of the $l$-dependent potential (dashed line) with the same quantities
calculated with the full $l$-dependent potential (solid line). The substantial
change in both quantities due to the $l$ dependence includes the appearance of a
conspicuous minimum near 140$^{\rm o}$.
\begin{figure} \caption{\label{cs1MK}  For 30.3 MeV protons on $^{40}$Ca,  the
solid lines are the angular distribution (above) and analyzing power for the
$l$-dependent potential of Ref.~\cite{kobmac79}. The dashed lines are calculated
with the same potentials except that the $l$-dependent component is omitted; the
difference represents the effect of the $l$-dependent component.} \begin{center}
\psfig{figure=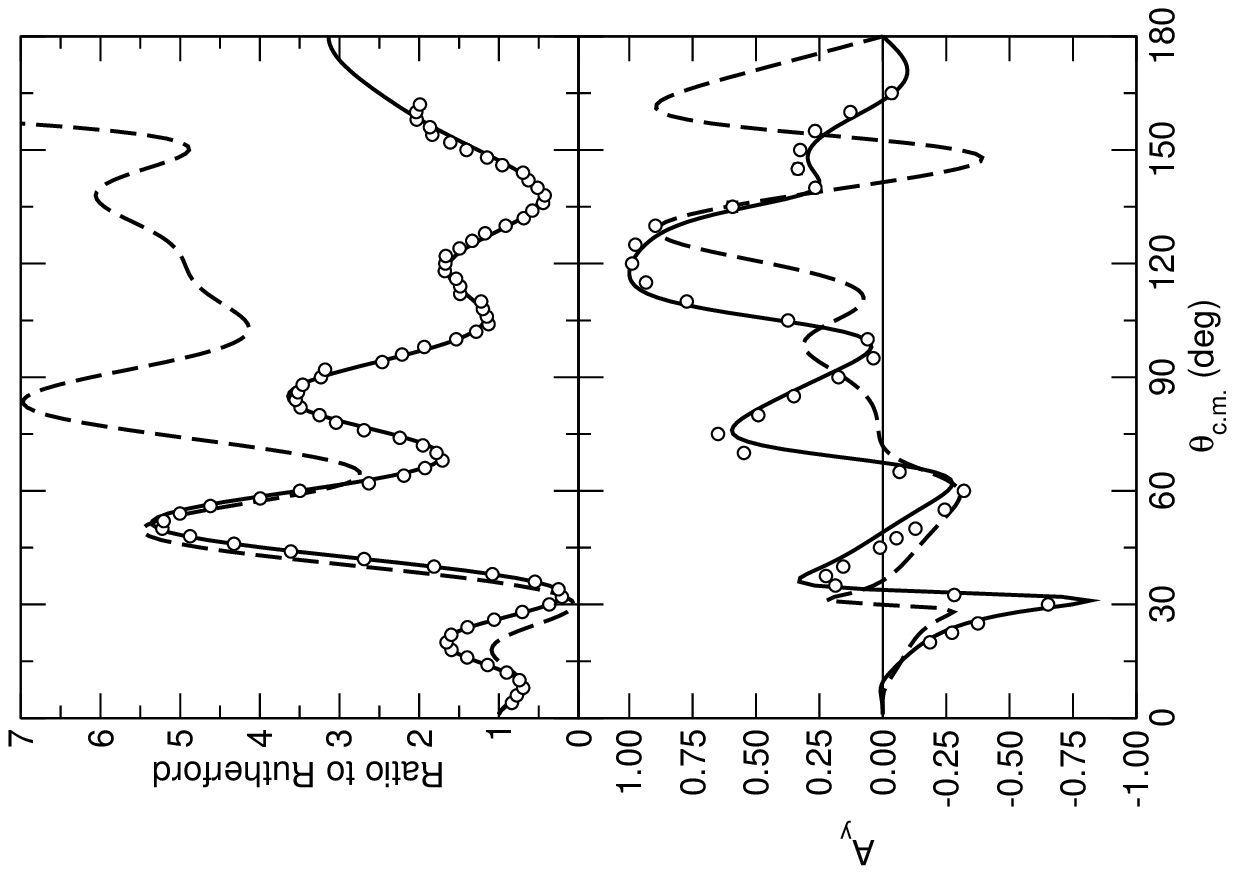,width=8cm,angle=-90,clip=} \end{center} \end{figure}

The $l$-dependent form of  Ref.~\cite{cordero} was applied in fits to elastic
scattering data for nucleons on  \nuc{16}{O}, \nuc{40}{Ca}, \nuc{56}{Fe} and \nuc{58}{Ni}
over a wide range of energies in Ref.~\cite{kobmac79} and applied to further
nuclei from \nuc{15}{N} to \nuc{208}{Pb} in Ref.~\cite{kobmac81}.
Good fits over a wide range of energies were found with parameters
and with properties (such as volume integrals and rms radii of the
$l$-independent component) that behaved in a more regular fashion than the same
properties of the best standard $l$-independent WS  fits. There were
suggestive exceptions in which resonance-like features appeared at certain
energies on otherwise smoothly varying volume integrals and rms radii. The same
quantities for the corresponding best $l$-independent WS fits were more irregular.
Ref.~\cite{kobmac79}, in which the contribution of the $l$-dependent terms
was very large, see Figure~\ref{cs1MK}, presents better fits for 30.3 MeV
protons on $^{40}$Ca  than Ref.~\cite{cordero}. Contributions like this
were part of a consistent pattern applying for a range of target nuclei and
energies. The larger effect evident in Figure~\ref{cs1MK} is a consequence
of the fact that the  $l$-independent term of Ref.~\cite{kobmac79}
was, for 30.3 MeV,  rather different from that of the earlier single-energy
fit of Ref.~\cite{cordero}.

The properties of the $l$-independent potentials that are  $S$-matrix-equivalent to the $l$-dependent
potentials of Ref.~\cite{kobmac79}, are significant.  For 30.3 MeV protons on \nuc{40}{Ca},
the undulations in both the real and imaginary central terms qualitatively resemble the undulations
generated by phonon coupling as presented  in Ref.~\cite{mk90} and discussed in Section~\ref{cplg}. In particular, the imaginary central term has excursions into emissivity, Ref.~\cite{available}, that have
magnitude and radius similar to those generated by coupling to phonons. These emissive regions do not
lead to the breaking of the unitarity limit.

\subsection{Evidence for $l$ dependence in \nuc{3}{He} scattering}\label{he3}
The angular distribution and analysing power for elastic scattering of \nuc{3}{He} at 33 MeV
from \nuc{58}{Ni} could not be fitted by standard Woods-Saxon  phenomenology. In
particular, the fit to the angular distribution was poor between 120 and near 180 degrees.
In Ref.~\cite{rm-he3}, the same data were analysed with the same form of $l$
dependence~\cite{cordero,kobmac79} that had been applied to proton scattering.
The qualitative fit  to the angular distributions was greatly improved for $\theta \geq 120$ degrees
with $\chi^2/N$ halved. Subsequently, the same $l$ dependent model was applied,
Ref.~\cite{brum-he3}, to the scattering of polarised \nuc{3}{He} on \nuc{16}{O} and
\nuc{40}{Ca} but in this case the $l$-dependent component did not improve the fit.
The significant difference was that for both the \nuc{16}{O} and \nuc{40}{Ca}
cases the data terminated below 120 degrees. This is clear example of a case in which
incompleteness of the data has concealed possible evidence for $l$-dependence. It is a
pity since \nuc{16}{O} and \nuc{40}{Ca} are, like \nuc{58}{Ni},  of low collectivity, leading to
elastic scattering angular distributions with well defined deep minima. Ref.~\cite{cage} presents
elastic scattering angular distributions (ADs) for 33 MeV \nuc{3}{He} on Ni isotopes of varying collectivity.
The backward angle fits for the least collective isotope, \nuc{58}{Ni}, with standard Woods-Saxon
potentials, are much poorer than the fits for the more collective isotopes. We know this because of
the relative completeness of the AD data for all these cases which extend to about 175 degrees.
This relationship between collectivity and ease of fitting applies to Ca isotopes for which  it was the least
collective isotope, \nuc{40}{Ca}, that revealed the requirement for $l$ dependence in nucleon scattering.
It appears that, for nuclei with low lying collective states, competing processes tend to wash out the
sharp features in the ADs that make $l$-dependence manifest.

\subsection{Interpretation of $l$ dependence found by fitting data}\label{interp}
Given a phenomenological $l$-dependent potential, two questions arise:
How is the $l$-dependence related to the coupling potentials due to inelastic and reaction
channels and evaluated by $S_l \rightarrow V(r)$ inversion?
What is the relationship to the wavy $l$-independent potentials found by
precise model-independent fits to the same data?

\subsubsection{Relating $l$ dependence to the effects of channel coupling.} \label{cplg}
The first question can be approached in two ways: firstly, directly compare the
changes in $|S_{lj}|$ and $\arg (S_{lj})$ that are due to channel
coupling with the change in these quantities arising from the $l$-dependent terms.
This was done in Ref.~\cite{mackob79} where  a
qualitative similarity was found. Alternatively,  invert $S_{lj}$ from the
$l$-dependent potential and compare the properties of the
resulting $l$-independent potential with the properties of potentials found by inverting
the elastic channel $S_{lj}$ from the coupled channel calculation.

\begin{figure}
\caption{\label{dpp1}  For 30.3 MeV protons on $^{40}$Ca, the four components of
the $l$-independent equivalent of the full $l$-dependent potential of
Ref.~\cite{cordero} (dashed lines) and of Ref.~\cite{kobmac79} (solid lines)
with, in each case, the respective $l$-independent part of that potential subtracted. From
the top: the real central, imaginary central, the real spin-orbit and imaginary
spin-orbit components.}
\begin{center}
\psfig{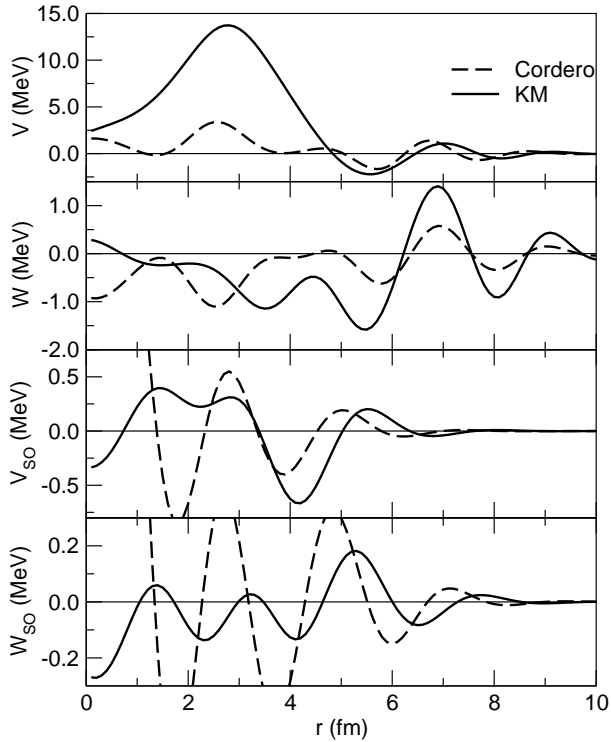} \end{center} \end{figure}

We first compare the two cases of Section~\ref{fits}, Ref.~\cite{cordero} and  Ref.~\cite{kobmac79},
subtracting from the (wavy) $l$ independent potentials, which were determined
by inverting $S_{lj}$ from the $l$-dependent potentials, the respective (smooth)
$l$-independent  component of the respective $l$-dependent potential. We associate the
wavy difference potential with the $l$ dependence. Figure~\ref{dpp1} presents the results
for the two cases: the dashed lines are for the $l$-dependent potential of Ref.~\cite{cordero}
(`Cordero' in the figure)  and the solid lines are similarly for Ref.~\cite{kobmac79} (`KM' in
the figure).  The differences between these reflect the superior fit to data by the KM potential
leading to the larger effect noted above. Since each parameter search fitted both
the $l$-independent and  $l$-dependent components, the curves do not involve subtraction
of the same  $l$-independent terms. Nevertheless,  there are common properties,
beyond the simple fact of undularity.  The surface region is significant, and the `wavelength'
of the undulations is the same in each case,  and, for the real central terms the amplitude
is greater for the KM case. A significant feature, as noted in Section~\ref{fits} above,  is
the existence of emissive regions in the imaginary central term near 7 fm and 9 fm. These persist
in the actual (unsubtracted) potentials. Such emissivity, having restricted radial range and not breaking
the unitarity limit, commonly arises both as a result of channel coupling and also in phenomenological
$l$ dependence, as here. For $r<5$ fm, the central terms for both the solid and dashed lines,
the Cordero and  KM cases, exhibit repulsion and absorption. This is also a general characteristic
of the coupling to deuteron  channels, as shown by the DPPs from the CRC calculations of
Ref.~\cite{pca40-2012,erratum,remark,tobepub}. These  pickup generated DPPs
also exhibit emissive regions in the central imaginary  term  at 7.5 fm (see Fig,~\ref{fig1})  and 9.5 fm (not
shown in Ref.~\cite{pca40-2012,erratum,remark,tobepub} for 9.5 fm.)  The KM and Cordero spin-orbit components
have a generally similar undularity, with repulsion around 6 fm in the real terms and emissivity
for both imaginary parts around 5 fm.

Although the DPPs from the pickup coupling calculations show similar undularity in the surface
region,  for $r <6$ fm, to that arising from $l$ dependence, the DPPs
for the central terms in Ref.~\cite{pca40-2012,erratum,remark,tobepub} and Fig.~\ref{fig1}
are relatively smooth, exhibiting repulsion in the real part and absorption in the imaginary part.
However, pickup channel coupling is not the only coupling that might lead to $l$ dependence.
Refs.~\cite{pig81,del86} cited previously demonstrated that coupling to giant resonances had
a significant effect. This effect was directly compared  in Ref.~\cite{del86} with the contribution to $S_{lj}$
of the phenomenological $l$-dependent term for just $|S_{lj}|$ but unfortunately  not for $\arg (S_{lj})$,
which is most directly related to the real part of the potential. In Ref.~\cite{mk90} the DPP due to coupling to
a large set of phonons for 30.3 MeV protons on \nuc{40}{Ca} had emissivity in the imaginary
part near 6 fm and 9 fm, and quite large amplitude undulations over the full radial range. As
mentioned in Section~\ref{fits}, the $l$-independent equivalent to the empirical $l$-dependent
potential does have emissive excursions in the  surface imaginary term that match the surface
emissivity, among other features, of this phonon-induced DPP quite well.

At present there have been no realistic calculations of the DPP including both reaction channels and
collective states together. It should become possible in future to make realistic comparisons
between the local $l$-independent  equivalents to both (i)  potentials derived from comprehensive
channel coupling, and, (ii) $l$-dependent potentials that precisely fit data. At present we can
only observe the qualitative similarities.

\subsubsection{Relating $l$-dependent and model-independent potentials.}\label{tuf}
Ref.~\cite{kmAnnP} presented $l$-independent potentials fitted to elastic
scattering angular distributions and analyzing powers for protons scattering
from \nuc{16}{O} and \nuc{40}{Ca} for various energies. These model independent
fits using spline functions were described as `theoretically unprejudiced fits'
although it is now clear that a prejudice was imposed: the prejudice that the
imaginary part of the potential should be absorptive everywhere. It is now
understood that this is not a necessary condition for $|S_{lj}| \le 1.0$ (the
unitarity limit) and oscillatory imaginary potentials can have localized
emissive regions without breaking the unitarity limit. Moreover, as mentioned
in Section~\ref{equiv}, the lack of suitable Wolfenstein (spin rotation) data makes
fully unambiguous theoretically unprejudiced fits formally impossible for proton scattering.
However, model independent fitting absolutely requires wavy potentials, and the
waviness found for the case of \nuc{40}{Ca} does share some features with that
in Fig.~\ref{dpp1}, in particular repulsion near 3 fm. Phenomenology based
on fits at a single energy is further complicated by the possibility of `transparent
potentials', i.e.\  (highly undulatory) potentials that, when added to another potential,
make effectively zero change to $S_l$ or $S_{lj}$.

\subsection{Model calculations linking $l$ dependence and undularity}\label{wavy}
The DPP generated by coupling to specific channels, as determined by inverting
the elastic channel S-matrix from the coupled channel calculation,  is generally
undulatory. For proton scattering, the DPP due to pickup coupling is
invariably rather wavy. This waviness is not an artefact of the inversion
procedure and is not restricted to proton scattering, but also applies to the
coupling to breakup channels for composite projectiles. For example, following
$^6$Li + $^{12}$C continuum discretized coupled channels, CDCC, breakup
calculations~\cite{pmPRC84}, there emerged a tendency for the local DPP due to
breakup of $^6$Li to be somewhat wavy in the surface for the lowest energy (90
MeV) case.  Calculations~\cite{mpPRC86} of deuteron breakup on \nuc{58}{Ni}
revealed that \mS\ often increases as a result of processes that might be
thought absorptive. When that study was extended down to
50 MeV, a quite  significantly wavy shape appeared in the surface of the
inverted potential. The undulations make a nearly zero contribution to the
volume integral to the potential. The wavy form of the DPP does not correspond
to the radial shape of the excitation or transfer form factor.

To get some understanding of these undulations, simple model calculations for
that case, 50 MeV deuterons on \nuc{58}{Ni}, were performed. A
basic question was posed:  given $S_l$ calculated from an $l$-independent potential, what
$l$-dependent modification of this $S_l$  might give rise to undularity of the
kind that has been found? This aspect of potential scattering theory has had
little attention. The argument, $\aS = 2\delta_l$, and modulus, \mS\, of the
S-matrix $S_l = \exp (\rri\ \aS) \mS$ calculated from a standard WS potential
were independently modified (\aS\ and \mS\ relate mostly to the real and
imaginary parts of the potential respectively) and the new S-matrix was
inverted. We here briefly describe results for modifications such that $S_l$ was
unchanged for lowest $l$ and either \mS\ or \aS\ was modified for high-$l$, with
a smooth transition; a fuller account is available in~\cite{preprint}. In both
cases the inverted potential had undulations. The undulations had a larger
amplitude in the real part when \aS\ was modified and a larger amplitude in the
imaginary part when \mS\ was modified. It is noteworthy that the modification of
\aS\ had a much larger effect on $J_{\rm R}$ than on $J_{\rm I}$ and effectively
\emph{zero} effect on the total cross section although the elastic scattering
angular distribution \emph{was} modified significantly. That is, \emph{a large
modification in the angular distribution was accompanied with essentially zero
change in the total reaction cross section.}

The modification of \mS\ was such that, $(1- \mS)$ was multiplied by
\begin{equation} f_{\rm m}(l) = 1 + z_{\rm m} \frac{1}{1 + \exp{((l -l_{\rm
m})/a_{\rm m})}} \end{equation} for $l_{\rm m}=14$, $z_{\rm m} =0.1$ and $a_{\rm
m} =2 $ with the asymptotic effect:\\ for $l \ll l_{\rm m}$, $|S_l| \rightarrow
|S_l|$,\\ for $l = l_{\rm m}$, $1- |S_l| \rightarrow (1 - |S_l|) + \frac{z_{\rm
m}}{2} (1 - |S_l|) ,$ and\\ for $l \gg l_{\rm m}$, we have  $1- |S_l|
\rightarrow (1 + z_{\rm m}) (1- |S_l| ).$

The effect of this was to increase the volume integral of the imaginary part
of the inverted potential, $J_{\rm I}$,  and increase the reaction cross section.
It also  induced Fraunhofer-like oscillations on the elastic scattering angular
distribution. The effect was linear insofar as, for example, all these effects
changed sign for $z_{\rm m} =-0.1$. When the modified $S_l$ was inverted,
the most relevant effects were: (i) very strong oscillations appeared
in the imaginary potential, (ii) oscillations also appeared in the real part
but these corresponded to very small changes in the volume integral $J_{\rm R}$
and rms radius, (iii) the oscillations in the imaginary part in the surface included excursions
into emissivity. There was no question of the unitarity limit being broken since
the modification of \mS\ did not allow that.

In Section~\ref{composite} we will associate strong undulations with a rapid change
in $S_l$ around the $l$ values for which $|S_l| \sim \hlf$, and that is supported by
the above results. Moreover, we should not expect waviness in just  the real
or the imaginary component.  Point (iii) is particularly significant, telling us  not to exclude,
on unitarity grounds, the occurrence in model independent fits of local radial regions
where the imaginary  component is emissive.

\subsection{Evidence for parity dependence from fits to data}\label{parityfits}
The generation of parity dependence by exchange processes  was recognized in
phenomenological analyses: calculations  for $ n + \alpha$ scattering~\cite{ttPRC4}
included a Majorana term in the real potential. Subsequently, an imaginary Majorana term
was included in an analysis of $p + \alpha$ scattering~\cite{ttbPRC5}. These
studies involving light target nuclei suggested the application to heavier
nuclei, and  real and imaginary Majorana terms were included~\cite{gmttPRC6} in an
analysis of proton scattering from $^{40}$Ca. This work showed that small
parity-dependent terms had a large effect at far backward angles but it was not conclusive
since other effects are clearly important in this case, as discussed in
Section~\ref{Ca40} below.  However, this work did inspire a more extensive
exploration~\cite{vos79} of the possible need for Majorana terms in the general
nucleon OMP. Ref.~\cite{vos79} found that Majorana terms were important for $p +
^{16}$O, less so for a $^{40}$Ca target and negligible for scattering from heavier nuclei.

Parity dependence has been firmly established by fitting experimental
data for nucleon scattering from \nuc{4}{He}~\cite{PRC54} and
\nuc{16}{O}~\cite{NPA618}  and  for \nuc{3}{He} scattering from \nuc{4}{He}, see
Ref.~\cite{npa713}. The parity dependence of the interaction for  \nuc{4}{He} scattering from
\nuc{12}{C}~\cite{NPA517} has been established by inverting $S_l$
precisely fitted to data over a range of energies. A potential applicable
at the lowest energies was determined. The volume integral $J_{\rm R}$
(defined in Ref.~\cite{satchler}) for odd parity was  $\sim425$ MeV fm$^3$
whereas that for even parity was $\sim390$  MeV fm$^3$. This difference might be
significant for  tunneling at astrophysical energies where $l=0$ dominates; in that context
a parity-independent potential fitted to data for this system would be
influenced by the need to fit the odd-parity $S_l$.

\subsection{The case of proton scattering from \nuc{16}{O}}\label{SGC}
Remarkably precise wide angular range data for angular distributions and analysing power
exist for proton scattering from \nuc{16}{O} from about 20 to 50 MeV.
Many attempts have been made to fit these, including spline fitting~\cite{kmAnnP}
in which precise fits led to undulatory potentials (but see comments in Section~\ref{tuf}).
The scattering data were very well fitted with an $l$-dependent potential~\cite{kobmac79}
the characteristics of which varied much more smoothly with energy than the characteristics
of the best fitting $l$-independent potentials.  For 30.1 MeV protons, comparing $l$-dependent
and $l$-independent fits, $\chi^2/N$  for the angular distribution was two orders of
magnitude lower for  the $l$-dependent fit and for the analysing power one order of magnitude
lower for the $l$-dependent fit. While far short of the $\chi^2/F \sim 1$ in principle achievable
with model independent fitting, the consistency of the potential over the range of  energies
and the vast superiority over conventional Woods-Saxon fitting, was conspicuous.

These results were obtained before the substantial parity dependence of the  proton-\nuc{16}{O}
interaction was established. This parity dependence was manifest in the inversion of $S_l$ from
RGM calculations  for protons up to 25 MeV, Ref.~\cite{PRC54}. The Majorana term of the inverted
potential was strongly repulsive for $r < 2 $ fm to an extent that was less at around 25 MeV
than at zero energy. The sign of the Majorana term is opposite to that for the proton-\nuc{4}{He}
interaction and, as  expected~\cite{BayeNPA460},  less in magnitude.

The RGM results are consistent with the most comprehensive fits to elastic scattering data:
a good fit to the high  quality angular distribution and analysing power data over a wide range
of energies,  was achieved by Cooper~\cite{NPA618} using direct data-to-potential
energy-dependent inversion. Single energy fits were also
found with $\chi^2/F$ values ranging from about 3 to about 9 for energies
from 27 to 43 MeV, far lower than for conventional phenomenology, although
greater than for precision single-energy model-independent fitting. From this work
there emerged a complex, parity-dependent potential, that was remarkably consistent
over the whole energy range studied.
Two features stand out:\\ (i) the real central Majorana term is repulsive for small $r$
and attractive further out, remarkably like the Majorana term found by inverting
the RGM $S_l$,  Ref.~\cite{PRC54}. \\
(ii) the imaginary central term is strongly emissive for $r < 2$ fm.\\
This latter feature would certainly require an $l$-dependency in order to be represented
by a conventional smooth potential.

For nucleons scattering from \nuc{16}{O}  the local equivalent
contributions of coupled collective states and also reaction channels (pickup channels)
have been determined separately and together~\cite{KMPRC97}.
Undulatory DPPs were found and there was a substantial difference in the pickup DPPs for protons
and neutrons, mostly due to different Q-values and transfer form factors for $(p,d)$ and $(n,d)$
processes. There were emissive regions in the DPPs.

From both theoretical and empirical evidence conclude that the proton plus \nuc{16}{O} potential certainly has both significant parity dependence, and also dynamical $l$ dependence. Any phenomenological treatment of
the proton plus \nuc{16}{O} system that ignores parity dependence is therefore deficient as is any treatment
ignoring the $l$-dependence arising from coupling to inelastic and transfer channels.

\subsection{The case of proton scattering from \nuc{40}{Ca}}\label{Ca40}
Section~\ref{cc} referred to the difficulty of fitting proton scattering from
\nuc{40}{Ca}.  This is an old problem:  in
a paper from 1967 by E.E. Gross \etal~\cite{eegross} we read:``The \nuc{40}{Ca} nucleus
has for some time been recognized as a b\^{e}te-noire  of  the optical model. [...]
It proved to be impossible to fit the scattering data beyond $140^{\circ}$ and
obtain a simultaneous fit to the polarization and cross-section data with a
reasonable set of optical-model parameters.'' Nearly 50 years later there is no
generally agreed solution to the problem of proton scattering from \nuc{40}{Ca}, but
there is now clear evidence that $l$ dependence is involved. Parity dependence is expected
to be much less for \nuc{40}{Ca} than for \nuc{16}{O}.

The $l$-dependent fit of Ref.~\cite{kobmac79} for 30.3 MeV protons
gave $\chi^2/N = 2.09$ for the angular distribution compared with the best
for a conventional Woods-Saxon fit of $\chi^2/N = 20.25$, with somewhat less
improvement for the analysing power.

Initially, coupling to deuteron channels appeared to
solve the problem of the backward angle dip~\cite{kobmac76}, but as the calculations
became more  complete this agreement disappeared. Coupling to highly excited resonance
states also improved the fit at backward angles to some degree~\cite{pca40-2012,erratum,tobepub}.
As noted in Section~\ref{survey}, coupling to a larger array of collective states made a large
contribution, but there is no full calculation that precisely fits the data. The calculations showed
the power of channel coupling to modify backward angle scattering, but they
are very incomplete and present a major parameter selection problem. What has emerged
however, is that the local equivalent DPP generated by coupling a plausible large collection
of collective states is notably undulatory, and, as remarked above, exhibits a large emissive
feature in the imaginary part for $r<2$ fm. The model independent phenomenological
fits of Ref.~\cite{kmAnnP} exhibited undulatory features but with a different wavelength.

\nuc{40}{Ca} that coupling to a reasonable array of collective states generates a dynamic
polarization potential that is strongly emissive for $r <2$ fm~\cite{mk90}.

Section~\ref{equiv} referenced a model independent fit to the 30.3 MeV data by Alarcon
\emph{et al}~\cite{ARF}. The deep  backward angle minimum was not fitted;
restrictions had been imposed on the model independent fitting to ensure that the
radial form was free of undulations.

A characteristic of the angular distributions for elastic proton scattering from \nuc{40}{Ca}
is the deep, hard to fit,  minimum.  This is probably related to the small
number of competing processes for  closed  shell \nuc{40}{Ca}.  If $l$ dependence is
established for \nuc{40}{Ca},  it should be assumed to be a general property,
even where it is easy to fit large angle elastic scattering data.

\subsection{Low energy neutron scattering}\label{lo-energy}
Low energy scattering of neutrons from the doubly closed shell  nucleus \nuc{208}{Pb} is
of special interest as a testing ground for potentials that are consistent with dispersion
relations over a wide energy range, as indicated in Section~\ref{causality}.  Studies of this
system have revealed evidence for $l$ dependence of a form distinct from those discussed above.
Such $l$-dependence appears first in Ref.~\cite{HJM,HJFMC}, a study of neutron elastic scattering
on \nuc{208}{Pb} for energies between 50 and 1005 keV. This work presents evidence for some
degree of $l$-dependence in the real part of the potential involving very few partial waves.
Specific resonant states play a role. Although suggestive, this is not evidence
for the kind of $l$ dependence implicit elsewhere.

A larger number of partial waves were involved in the wide energy range study of n + \nuc{208}{Pb}
elastic scattering~\cite{JHM}. All the data were fitted to a satisfactory standard with a potential model
that was consistent with causality and which had fixed geometric parameters. This review cannot do
justice to their work except to note that, at the lowest energies, they did find it necessary to apply a
form of $l$-dependence which was different from those presented elsewhere in this article. This
$l$ dependence applies in a regime where relatively few partial waves are involved. The partial
waves $l =0$ to 6 were divided into two groups, group $b$ containing $l = 1, 3, $ and 6, with the
remainder in group $c$; this is clearly not parity dependence. The imaginary surface potential was
different for groups $b$ and $c$ with the dispersion relations introducing a corresponding $l$
dependence for the real surface potential. The $l$ dependence, which  applies below 12 MeV and
 down to negative energies, does improve the fit to data in that energy range.

A later paper, Ref.~\cite{JW}, applied a similar idea but based on a somewhat different grouping of
partial waves. In this case the grouping was based on the different relationship between the radial
position of the antinodes of the partial waves and the  maximum magnitude of the surface imaginary
potential. This is clearly reasonable, but depends upon the certainty with which radial parameters
of the imaginary part are determined. There may well be a higher order effect in that coupling to
inelastic channels, which in a macroscopic picture involves derivatives of the potentials, will be
dependent on the form of the imaginary potential. Furthermore, in situations where there are few
operative partial waves, the coupling effects for individual partial waves are less likely to average out.

The $l$-dependent model  of Ref.~\cite{JHM} was further developed in Ref.~\cite{JJM} for
n + \nuc{208}{Pb}  elastic scattering for  energies between $-20$ MeV and 20 MeV. In this case
the $l$ dependence allowed good fits when the radial form of the potentials were energy independent.
The authors note that a dependency of the imaginary part upon angular momentum is equivalent to a
form of non-locality that is distinct from that due to exchange, citing~\cite{rawit87}. At this point contact
is made with a recurring theme of the present article.

It is hard to relate these low energy cases, involving specific partial waves that are related to specific bound
nucleon orbitals, to the forms of $l$ dependence involving many partial waves and related to the radial
shape of the target nucleus. As with all $l$ dependence, an alternative representation of the elastic scattering
involving an undulatory $l$-independent potential is possible, but this would not be equivalent when applied in reactions
involving the wave function in the nuclear interior.

\section{Scattering of heavier nuclei}\label{composite}
There is substantial literature concerning
$l$ dependence in heavy-ion scattering with independent
arguments for $l$ dependence of the real and imaginary components.
Results given in sections~\ref{HI-antisymm} and~\ref{reduced} suggest that when
the real and imaginary parts of a potential have different sources of $l$ dependence,
the consequences of the $l$ dependence of the real and imaginary terms
persist in the complete $l$-independent potential found by inversion.
In section~\ref{strong}, $l$ dependence due to strong inelastic coupling is discussed for heavier nuclei.

\subsection{$l$ dependence due to antisymmetrization}\label{HI-antisymm}
An example of $l$ dependence in the real part is provided by the RGM calculations
of Wada and Horiuchi~\cite{WH} for \nuc{16}{O} + \nuc{16}{O} elastic scattering.
The $l$ dependence arises from exchange terms that go far beyond the 1-particle
knock-on exchange. Horiuchi~\cite{horiuchi} reviews such calculations in the
context of a more general discussion of microscopic nucleus-nucleus potentials. There is no
possibility in this case of there being Majorana terms, although such terms will arise when
the interacting nuclei are not identical bosons. The $S_l$
corresponding to the $l$-dependent real potentials of Wada and Horiuchi were
inverted~\cite{ATMCW} to yield an $l$-independent potential which is
significantly different at lower energies from that derived~\cite{WH} using WKB
methods. The difference between the complete $l$-independent equivalent
potential and the $l$-independent (non-exchange) part of the Ref.~\cite{WH}
potential is most marked in the nuclear interior. Such $l$ dependence would
be  less significant for a potential that includes an absorptive term. Nevertheless,
Ref.~\cite{WH} established that exchange processes lead to an $l$ dependence
of nucleus-nucleus interactions that is in addition both to possible
parity-dependence and also  to contributions from knock-on exchange.

The modelof Kondo \emph{et al}~\cite{KRS}, for \nuc{16}{O} + \nuc{16}{O} scattering over
a range of energies, included a phenomenological $l$-dependent real term inspired by
the model of Wada and Horiuchi, together with an $l$-dependent imaginary term of
the form discussed in Section~\ref{reduced} below. The $S_l$ for the potential with
both terms $l$-dependent was inverted~\cite{ATCM} leading to a real potential
with a very similar shape and energy dependence to that found~\cite{ATMCW} for
the Wada-Horiuchi potential.

The $l$ dependence of the real part of the Kondo \emph{et al}~\cite{KRS}
potential was an overall factor $V_0 + V_l l(l+1)$, i.e. a more gradual $l$
dependence than the sharp transition involving a Fermi form employed
elsewhere: $1/\{1+ \exp[(l-l_{\rm c} )/\Delta]\}$. This, by
design, leads to a very similar energy dependence for the $l$-independent
potential found by inverting the  Wada and Horiuchi~\cite{WH} S-matrix.
The imaginary part of the potential was of Fermi form as in Section~\ref{reduced} below,
but in the 59 MeV case,  $l_{\rm c}$ was 19, rather higher than $k R_{\emph W} = 12.7$
where $R_{\emph W} $ is the effective radius defined by Kondo \emph{et al}. In
test cases, it has been found that a sharp transition in the potential
for $l$ close to the value for which $|S_l|\sim 0.5$ consistently leads to
an undulatory equivalent potential, generally more undulatory than those
found~\cite{ATCM} from the S-matrix of  Ref.~\cite{KRS}.
That exemplifies  an apparent systematic qualitative difference between the equivalent
$l$-independent potentials found for these `gradual' $l$-dependencies and the
sharper Fermi-form $l$-dependencies. The latter, especially for sharp changes in
the imaginary term, generate more oscillatory equivalent potentials.

The $l$-independent potentials $S_l$-equivalent to the Kondo potential  had substantially
non-zero radial gradients at the nuclear centre and very different wave functions in the overlap
region.

\subsection{$l$ dependence due to reduced absorption for high-$l$ partial waves}\label{reduced}
Following arguments
of Chatwin, Eck, Robson and Richter (CERR)~\cite{cerr70}, explicit angular momentum dependence
was introduced into the imaginary part of the OMP for heavier ions such that the absorptive term was
reduced  for the highest partial waves. This was justified, with reference to
Feshbach's theory, on the grounds of the  reduced number of channels for
available for absorption for these partial waves.

This CERR approach has had some success and has been applied in various cases,  not all
confined to heavier ions. The first applications involving alpha particles
gave consistent improvement to $\alpha$-particle elastic scattering below 20 MeV~\cite{bisson1}.
A CERR term was incorporated in the analysis by Bisson \etal~\cite{bisson2} of $\alpha$ scattering
from \nuc{40}{Ca} in a study in which compound elastic scattering also played a key role.
In this particular implementation,  the imaginary term was a standard Woods-Saxon
derivative form multiplied by the $l$-dependency:
\beq
F(l) = \frac{1}{1 + \exp ((l-l_c)/\Delta_l)} \label{J-l-dep}
\eeq
where, following CERR, $l_c =\bar{R} \times [E_{\rm{CM}} + \bar{Q}]^{\hlf}$.
In this case $\bar{Q} =0$ and $\bar{R}$ was chosen close to the interaction radius.
The high $l$  cutoff was quite broad, with $\Delta_l =4.0$. Ref.~\cite{bisson2} implies
that this $l$ dependence, together with an incoherent compound elastic component,
was essential for achieving a reasonable fit over the  energy range, 5.5 to 17.5 MeV.

More often, a CERR term has been included for heavier ion scattering such as
the model of Kondo \emph{et al}~\cite{KRS}.
Inversion, Ref.~\cite{ATCM}, reveals that except at the highest energy, the
$l$-independent equivalent of the imaginary part has a radial form
radically different from that of any $l$-independent potential found by fitting
data. The  CERR term was included together
with a parity-dependent real potential for \nuc{16}{O} + \nuc{20}{Ne} scattering
by Gao and He~\cite{gaohe} and the resulting $S_l$ were inverted~\cite{ATMC} to
produce an  $l$-independent representation. The resulting imaginary potential
was qualitatively similar to that produced~\cite{ATCM} by the model of Kondo
\emph{et al}~\cite{KRS}.

In Ref.~\cite{bohn},  CERR $l$-dependence applied to \nuc{6}{Li} scattering
from \nuc{40}{Ca} led to an improvement to the fit at backward angles. Ref.~\cite{bohn}
included a comparison of the number of exit channels as a function of exit channel
angular momentum for scattering from  \nuc{44}{Ca}. The result supports the apparent requirement for
CERR $l$-dependence for  \nuc{6}{Li} scattering from closed shell nuclei. For such nuclei the fewer
high-$l$ exit channels is in line with the basic CERR hypothesis. In fact, the formulation of
CERR~\cite{cerr70} was in terms of the conserved quantity $J$, the total angular momentum.
For zero target and projectile spins, $J =l$, but for the scattering of \nuc{6}{Li} the difference
is significant, especially where vector analysing powers are to be fitted, as in  Ref.~\cite{trcka}.
In that case the fit to the analysing powers was improved by the inclusion of what, in
this case, was $J$ dependence following Eq.~\ref{J-l-dep} applied to $J$ rather than $l$.
Another example is Ref.~\cite{Retal}, in which $J$-dependent absorption was included together with a
tensor interaction to fit a full set of tensor, as well as vector, analysing powers
in polarized \nuc{6}{Li} scattering from \nuc{12}{C}.

Ref.~\cite{ATMR} shows how the energy dependence of
the CERR $l$-dependent cutoff leads, by way of dispersion relations,
to an $l$ dependence in the real potential for \nuc{16}{O} + \nuc{16}{O} scattering.
Although there have been successful applications of CERR  $l$ dependence, it
appears not to have become generally established for heavy ion or $\alpha$ scattering.

\subsection{Strong channel coupling  in \nuc{16}{O} scattering on \nuc{12}{C} at
330 MeV and 116 MeV} \label{strong}
Channel coupling induces DPPs in the interaction between heavy ions that have
similar features  to the DPPs arising in the scattering of nucleons and other
light ions.  This suggests an underlying $l$ dependence. We now present
evidence for this in the case of \nuc{16}{O} scattering on \nuc{12}{C} at
330 MeV and 116 MeV. Ohkubo and Hirabayashi~\cite{OH1} showed how, for 330 MeV
\nuc{16}{O} scattering from \nuc{12}{C}, the excitation of strongly excited
states in both nuclei greatly modified the elastic scattering angular
distribution in a way that explained some long-standing paradoxical features.
Subsequently, the elastic channel $S_l$ from the coupled channel calculations
were inverted to reveal DPPs that had quite strong and well-established undulatory
features~\cite{3author}.  The  possibility that this represents an underlying $l$-dependent
potential was not explored but it is likely, especially since there was no
apparent relationship between the undulations in the imaginary term and the radial
dependence of the form factors for the inelastic coupling.

Subsequently, Ohkubo and  Hirabayashi~\cite{OH2} performed similar calculations
on the same pair of nuclei at the much lower  energy of 115.9 MeV (resulting
in very interesting conclusions concerning rainbow scattering).  The elastic
channel $S_l$ has been inverted and very strong undulations have been
found~\cite{116}. The amplitude of the undulations, together with the smaller
number of partial waves at 116 MeV compared with 330 MeV made it impossible to
establish a unique inverted potential. Apparently there are too few partial waves
in this case to avoid the transparent potentials mentioned in Section~\ref{tuf}.

The undulations exhibited by the
alternative inverted potentials shared strong family characteristics. To
understand these, model calculations were carried out at 115.9 MeV  in which a standard
potential similar to the bare potential of Ref~\cite{OH2} was made artificially
$l$-dependent and $S_l$ from this $l$-dependent potential was inverted.
The imposed $l$ dependence was simple and in the form of added terms $v(r)
\times f(l)$  or $w(r) \times f(l)$ where the $f(l)$ factor multiplying a real
($v(r)$) or  imaginary  ($w(r)$) terms is given by: \begin{equation} f(l) =
\frac{1}{1 + \exp{((l^2 -{\cal L }^2)/ \Delta^2)}} \label{lfac}. \end{equation}
The $v(r)$ and $w(r)$ factors each have a Woods-Saxon form with
the same radius and diffusivity parameters as the corresponding real and
imaginary $l$-independent terms. As a result,  the $l$-dependent potentials
essentially have a real or imaginary component that is
renormalized for $l$ less than $\cal{L}$, with a fairly sharp transition
since $\Delta$ is quite small. The potential is unmodified for   $l$ substantially greater
than $\cal{L}$.  The value of  $\cal{L}$ was chosen close the value of $l$ for which
$|S_l| \sim \hlf$. This sharp transition gives a form of $l$ dependence very different from
the more gradual form of Refs.~\cite{WH,KRS} of Section~\ref{HI-antisymm}. We
would not expect such a simple parameterization to exactly reproduce the
specific undulations, yet many of the general features, such as strong
undulations in the surface region of the imaginary term, emerged. This shows
again that it is quite possible to have a potential with distinct excursions
into emissivity but for which $|S_l| \le 1$ for all $l$, conforming to
the unitarity limit.

Details are presented in Ref~\cite{116}, but the conclusion is clear:
strong coupling to  states of both \nuc{16}{O} and \nuc{12}{C} induces a
DPP with an $l$-independent  representation having strong undulations.
Thus, a representation in terms of smooth potentials must have significant
$l$ dependence. The nature of this $l$ dependence is plausibly of a form
distinguishing between partial waves above and below the region where $|S_l|
\sim \hlf$.

\subsection{More general $l$-dependence in \nuc{16}{O} scattering}\label{generalO16}
Since the excitation of cluster states contributes to the scattering of \nuc{16}{O} from
\nuc{12}{C},  it must be presumed to play some part in scattering from heavier target nuclei.
The scattering of \nuc{16}{O} from \nuc{28}{Si} at about 55 MeV exhibited enhanced
backward angle scattering for which there has been no widely agreed explanation,
see Ref.~\cite{lee} for references.  In Ref.~\cite{MKsplines} the angular distribution
was  fitted with model independent searching using spline functions.
In the region of the strong absorption radius (SAR), about 9 fm, the spline fit agreed well
with potentials of standard parameterized form cited in Ref~\cite{lee}.  However, the closer
fit to the data found by the spline model led to a potential that deviated markedly
from the other fits in the range 6 - 8 fm, a range still important for a precise fit. This deviation
was, in effect, part of a strong undulation that was undefined at smaller internuclear separation.
Subsequent spline function fits~\cite{KSM} for \nuc{16}{O}  -- \nuc{12}{C} elastic scattering
from 33 to 55 MeV, consistently revealed similar marked deviations from folding model potentials.

These results bring into focus the choice of representation: $l$-dependence or wavy
potential? Possibly  there is parity dependence due to multiple alpha cluster
exchange, but it is unlikely that the undulations arise
from corrections to local density folding models. The wide angular range data does not yet
have an agreed explanation in terms of reaction dynamics, but it is certain that $l$-independent
smooth potentials are excluded.

There is indirect evidence for $l$ dependence applying to \nuc{16}{O} scattering
from the heavier \nuc{40}{Ca} target at $E_{\rm c.m.} = 37.5$ MeV/  It points to the need
for care in interpreting spline model fits,  see Ref.~\cite{KMPRC26}. Spline model
fitting revealed small amplitude ( $ \sim \pm 2$ MeV) undulations in the radial range
a few fm  within the SAR.  These were well-determined,
unlike the wide amplitude undulations of previous spline model fits cited in  Ref.~\cite{KMPRC26}.
A natural explanation is $l$ dependence in the underlying potential.
More precise angular distribution measurements would enable modern fitting
techniques to make a more definitive  determination.

\section{Implications and applications}\label{imp-app}
The $l$ dependence of the OMP is of intrinsic interest, but how is it to be taken into account
in reaction calculations?

\subsection{Practical implementation of $l$-dependence}\label{practical}
The possibility of $l$ dependence is an inconvenience. It is not
commonly an option offered by standard reaction codes that involve the application of
optical potentials. One problem is the wide variety of possible $l$ dependencies.
Nevertheless, there are certain forms of $l$ dependence that should certainly be available for
use in reaction calculations.
\begin{enumerate}
\item The interaction of light ions with lower mass targets will, in general, be parity dependent.
For example, for nucleon scattering on \nuc{16}{O}, or lighter, parity dependence should
not be omitted and it is very large for a $\nuc{4}{He}$ target.
\item The success of the CERR $l$ (or $J$) dependence for $\alpha$ or
\nuc{6}{Li} scattering implies that this form should be available for reaction calculations
over the appropriate energy range.
\end{enumerate}

Concerning point 1: For  $\alpha$ - \nuc{12}{C} scattering, it is the even parity potential
that is relevant to the astrophysical S-factor, but  a potential fitted $l$-independently will have been
influenced by odd-parity partial waves~\cite{NPA517}.  Where exchange processes lead
to parity dependence,  such dependence can be omitted only when exchange is included
explicitly.  Section~\ref{HI-antisymm} referred to  various other forms of $l$
dependence arising from antisymmetrization.

Concerning point 2: If  CERR $l$ or $J$ dependence were firmly established,
that would be a significant extension of the optical model. If  the plausible
formal arguments for CERR, in Ref.~\cite{cerr70},  were proven invalid, that would present an interesting challenge.

The other forms of angular momentum dependence which, together with dynamical non-locality,
arise from channel coupling,
present a problem: there exists no widely accepted parameterized form for inclusion in reaction
codes. In principle, it would be possible to include the many processes which generate $l$ dependence
within the direct reaction of interest. It is often stated, as in Ref.~\cite{OHO}, that there exist
elastic scattering angular distributions that cannot be described by a mean field optical potential.
But the existence  of cases where
smooth mean field OMPs do not work does not mean that a potential model fails. In cases like that of
Ref.~\cite{OHO}, angular distributions can be fitted when strong channel coupling is included. By
means of  S-matrix inversion, such coupling effects can always be represented
within a potential model. The potential will probably be undulatory, implying the existence of
an alternative representation in terms of $l$ dependence, although the form of the $l$ dependence
might not be easy to identify. In Ref.~\cite{OHO} the effect of coupling is very large,
and of great interest;  this is an extreme case of a general property.  In such cases there
is unlikely to be any reason to input the $l$ dependence in reaction codes for  calculations
of other processes.

\subsection{Consequences of $l$ dependence for folding models}\label{heavy}
Single folding calculations, based on theoretical nucleon
potentials of the kind discussed in Section~\ref{self}, have been applied with
some success~\cite{Pang,pang2} to the scattering of lighter composite nuclei.
It is unclear how an $l$-dependent nucleon potential should be incorporated in such 
single folding calculations. To the extent that $l$ dependence can be associated with 
calculable reaction processes, those reaction processes should be incorporated into 
the scattering calculations for the composite nuclei. For example, if the coupling to 
giant resonance states of the target is a major source of $l$ dependence for nucleon 
scattering, then the same processes must be presumed to affect the scattering of composite 
nuclei.  More generally, processes that lead to $l$ dependence for proton scattering 
presumably give rise to $l$ dependence for composite projectiles. This might be 
significant for lighter nuclei that are sensitive to more than the nuclear surface.

\subsection{Application of $l$-dependence in direct reactions}\label {direct}

The $l$-dependent extension of the nucleon OMP is relevant to the analysis of direct reactions.
Ref.~\cite{kobmac81} compares the angular distributions for the proton inelastic scattering
to the $3^-$ state of  \nuc{16}{O} calculated with both $l$-dependent and $l$-independent
OMPs. There is a considerable difference in the angular distribution away from the maximum.

There is a problematic aspect concerning the application of $l$-dependent potentials
in CRC calculations.  It is likely that a major source of $l$
dependence is the coupling of collective states or reaction channels to the elastic channel.
While it is interesting to study the application of  such potentials in CRC calculations,
it is inappropriate to include the same channel coupling, that contributes to the $l$
dependence, together with the explicit inclusion of that $l$ dependence, within a larger coupled
channel calculation. This is one aspect part of a general  non-trivial question, that we do not
pursue here, relating to the application of a potential, defined to reproduce
elastic scattering,  to other reactions, including fusion.

\section{Conclusions and discussion}\label{conc}
Interaction potentials between nuclei scattering from each other depend upon the
orbital angular momentum $l$ of their relative motion: that is the conclusion of
the results assembled here.
There are several distinct forms of $l$ dependence for which there are both different degrees of
certainty and different implications. It can be considered certain that the interaction between
nucleons and \nuc{4}{He} and even \nuc{16}{O} and also, for example, between \nuc{3}{He} and
\nuc{4}{He} are parity dependent, and this should be taken into account in analyses of these
cases. The  $l$ dependence of the imaginary potential of the CERR form
has not been widely adopted. It would be a genuine contribution to
our understanding of heavy ion interactions if the process behind CERR $l$ dependence
were firmly established or shown to be absent. In another category is the dynamically induced $l$ dependence
of the nucleon-nucleus  potential. There are both phenomenological and theoretical
arguments for this, and these arguments deserve to be either strengthened or disproved. The
nucleon nucleus interaction has a special status as being a positive energy continuation
of the shell model potential, and also being a vital ingredient in the
analysis of direct reactions, a subject of continuing
interest~\cite{nunes}. Dynamically induced $l$  dependence would have the status
of a generic phenomenon if that between interpenetrating heavier ions such as
\nuc{12}{C} and \nuc{16}{O}, as in Section~\ref{strong}, were firmly established.
Precision fits to  data will, in general (where the angular distributions are not too smooth)

reveal the need for a departure from local density model potentials. Only an unnecessarily
restrictive  form of OMP fails. What is  missing is a `dictionary' for interpreting undulatory
potentials in terms of specific $l$ dependencies.

Precise and complete elastic scattering data can always be fitted; model independent fitting
takes us from the situation where  particular data cannot be fitted well to the situation
where the same  data is manifestly incomplete. In fact, as exemplified in the \nuc{3}{He} case
discussed in Section~\ref{he3}, the incompleteness of existing data is the major
barrier to establishing  phenomenologically that $l$ dependence is a general property
of nuclear interactions.  Relatively complete high quality data exists for some cases
so it is a shame that the full information contained of such data is rarely
fully exploited in a systematic way. There is an understandable reluctance for  `just
fitting data', especially when such fits lead to strong undulations, Refs~\cite{ermer,ermer2}.
It is fortunate that Kepler did not feel
that way about fitting Tycho Brahe's high quality planet-Sun scattering data.

The thrust of the present work is that there is information concerning nuclear reactions, that is seldom
fully exploited,  that could be extracted from  elastic scattering data.  In fact the
data is seldom complete and absence of spin-rotation nucleon scattering data is a real problem~\cite{kmr}.
Nevertheless, we know that the success of conventional folding models to fit existing data is
incomplete as are present attempts~\cite{pca40-2012} to reproduce the data with
channel coupling effects. This suggests limits to the local density approximation.
Concerning CRC calculations, it is well established that the DPPs representing channel coupling  are never
proportional, as a function of $r$, to the bare potential. It follows that, by  approximately improving the fit
of a folding model potential to  data,  by means of a uniform renormalization, an opportunity
to extract information about reaction dynamics is lost. A model
independent additive term might well contain indications of $l$ dependence and be
identifiable with calculable DPPs.

There are some firm theoretical predictions for $l$ dependence for the
scattering of heavier nuclei, as we noted in Section~\ref{HI-antisymm}. For the scattering of nucleons
and certain light nuclei, there are \emph{direct} predictions of parity dependence
that are supported by experiment. The theoretical arguments for more general $l$ dependence
are less direct, apart from the relationship established between undularity and reaction coupling.
Establishing more direct evidence remains  a challenge.  It should not
be forgotten that it is when our favorite folding model \emph{fails} to give precise
fits that we stand to learn. Arguably, our understanding of
nucleon-nucleus scattering is incomplete even at the most phenomenological level.

\section{Acknowledgment} I am grateful to Nicholas Keeley for producing publishable figures
and for many helpful discussions.


\begin{thebibliography}{99}
\bibitem{darri}P. Darriulat, G. Igo, H.G. Pugh, and H.D. Holmgren, Phys.\ Rev.\ {\bf 137}, B 315 (1965).
\bibitem{relation}R.S. Mackintosh, arXiv:1705.07003v6 (2019).
\bibitem{inv1}R.S. Mackintosh and A.M. Kobos, Phys.\  Lett.\  {\bf B 116}, 95 (1982).
\bibitem{inv2}A.A. Ioannides and R.S. Mackintosh, Nucl.\ Phys.\ {\bf A 438}, 354 (1985).
\bibitem{inv3}A.A. Ioannides and R.S. Mackintosh, Nucl.\ Phys.\ {\bf A 467}, 482 (1987).

\bibitem{s-pedia}R.S. Mackintosh, Scholarpedia `Inverse scattering: applications in
nuclear physics', (2012).
\bibitem{Kuk04}V. I. Kukulin and R. S. Mackintosh, J. Phys. G: Nucl. Part. Phys. {\bf 30}, R1 (2004).


\bibitem{CM89}S.G. Cooper and R.S. Mackintosh, Inverse Problems {\bf 5},  707 (1989).
\bibitem{arxiv}R.S. Mackintosh, arXiv:1205.0468 (2012).

\bibitem{KD}A.J. Koning and J.P. Delaroche, Nucl.\ Phys.\ {\bf A 713}, 231 (2003).
\bibitem{DCV}W.W. Daehnick, J.D. Childs, and Z. Vrcelj, Phys.\ Rev.\ C {\bf  21}, 2253 (19780.
\bibitem{Pang}D.Y. Pang, P. Roussel-Chomaz, H. Savajols, R.L. Varner, and R. Wolski,
Phys.\ Rev.\ C{\bf 79}, 024615 (2009).
\bibitem{furu}T. Furumoto, K. Tsubakihara, S. Ebata, and W. Horiuchi, Phys.\ Rev.\, C{\bf  99}, 034605 (2019).


\bibitem{epja}R.S. Mackintosh, Eur.\   Phys.\  J.\ {\bf A 53}, 66 (2017).


\bibitem{FPW}H. Feshbach, C.E. Porter, and V.F. Weisskopf, Phys.\ Rev.\  {\bf 96}, 448 (1954).
\bibitem{WS}R.D. Woods and D.S. Saxon, Phys.\ Rev.\  {\bf 95}, 577 (1954).
\bibitem{gww}L.C. Gomes, J.D. Walecka,  and V.F. Weisskopf, Ann.\  Phys.\  (New York) {\bf 3}, 241 (1958).

\bibitem{feshbach}H. Feshbach, Ann.\ Phys.\ {\bf 5}, 357 (1958).
\bibitem{feshbach1}H. Feshbach,  Ann.\ Phys.\ {\bf 19}, 287 (1962).
\bibitem{bellsq}J.S. Bell and E.J. Squires, Phys.\ Rev.\ Lett.\ {\bf 3}, 96 (1959).

\bibitem{JLM}J.P. Jeukenne, A, Lejeune, and C. Mahaux, Phys.\ Rev.\ C{\bf 10}, 1391 (1974);
Phys.\ Rev.\ C{\bf 15}, 10 (1977)
\bibitem{MS}C. Mahaux and R. Sartor, Advances in Nuclear Physics, vol. 20, ed
J.W. Negele and E. Vogt (Plenum, New York, 1991), p. 1.

\bibitem{BR}F.A. Brieva and J.R. Rook, Nucl.\ Phys.\ {\bf A 291}, 299 (1977); Nucl. Phys.
{\bf A 291}, 317 (1977); Nucl. Phys. {\bf A 297}, 206 (1978).

\bibitem{temporal}C. Mahaux and G.R. Satchler, Nucl.\ Phys.\ {\bf A 560}, 5 (1993).

\bibitem{WT}K. Wildermuth and Y.C. Tang, \emph{A Unified Theory of the Nucleus} (Vieweg,
Braunschweig, 1977).

\bibitem{DD}P. Descouvement and M. Dufour, in C. Beck (ed.) \emph{Clusters in Nuclei Vol.2},
Lecture Notes in Physics, vol 848, p. 1 (2012).
\bibitem{SLYV}Y. Suzuki, R.G. Lovas, K. Yabana, and K. Varga, \emph{Structure and Reactions of Light Exotic Nuclei} (Taylor and Francis, London, 2003).

\bibitem{rao}C.L. Rao, M. Reeves and G.R. Satchler, Nucl. Phys. {\bf A207}, 182 (1973).
\bibitem{coulter}C.A. Coulter and G.R. Satchler, Nucl.\ Phys.\ {\bf A 293}, 269 (1977).
\bibitem{rawit87}G.H. Rawitscher, Nucl. Phys. {\bf A475}, 519 (1987).
\bibitem{ghj}N.K. Glendenning, D.L. Hendrie, and O.N. Jarvis, Phys.\ Lett.\ {\bf  26B}, 131 (1968).
\bibitem{rsm71}R.S. Mackintosh, Nucl.\ Phys.\ {\bf A 164}, 398 (1971).

\bibitem{satchler}G.R. Satchler, \emph{Direct Nuclear Reactions} (Clarendon Press, Oxford, 1983).
\bibitem{mkPRC81}R.S. Mackintosh and N. Keeley,  Phys.\ Rev.\, C{\bf  81}, 034612 (2010).
\bibitem{pca40-2012}R.S. Mackintosh and N. Keeley,  Phys.\ Rev.\, C{\bf  85}, 064603 (2012).
\bibitem{erratum}R.S. Mackintosh and N. Keeley,  Phys.\ Rev.\, C{\bf  98}, 069901(E) (2018).

\bibitem{mk90}R.S. Mackintosh and N. Keeley, Phys.\ Rev.\, C{\bf 90}, 044601 (2014).

\bibitem{jlmb1}E. Bauge, J.P. Delaroche, and M. Girod, Phys.\ Rev.\ C{\bf 58},  1118 (1998).
\bibitem{jlmb2}E. Bauge, J.P. Delaroche, and M. Girod, Phys.\ Rev.\ C{\bf 63}, 024607 (2001).
\bibitem{VM}N. Vinh Mau, \emph{Microscopic Optical Potentials}, Lecture Notes in Physics,
ed. H. V. von Geramb (Springer Verlag, New York, 1979), {\bf 89}, p. 104.
\bibitem{vmb}N. Vinh Mau and A. Boussy, Nucl.\ Phys.\ {\bf A 257}, 189 (1976).
\bibitem{owm}F. Osterfeld, J. Wambach, and V.A.Madsen, Phys.\ Rev.\ C{\bf 23},  179 (1981).
\bibitem{wmc}J. Wambach, V.K. Mishra, and Li Chu-Hsia, Nucl.\ Phys.\ {\bf A 380}, 285 (1982).
\bibitem{mo} V.A. Madsen and P. Osterfeld, Phys.\ Rev.\ C{\bf 39},  1215 (1989).
\bibitem{MW}C. Mahaux and H. Weidenm\"uller, \emph{Shell-Model Approach to Nuclear Reactions}
(North Holland Publishing Co., Amsterdam, 1969).

\bibitem{pig81}M. Pignanelli, H. V. von Geramb, and R. DeLeo, Phys.\ Rev.\ C{\bf 24},  369 (1981).
\bibitem{del86}J.P. Delaroche, M.S. Islam, and R.W. Finlay,  Phys.\ Rev.\ C{\bf 33}, 1826  (1986).
\bibitem{honorePRC33}G.M. Honor\'e, W. Tornow, C.R. Howell, R.S. Pedroni, R.C.
Byrd, R.L. Walter, and J.P. Delaroche, Phys.\ Rev.\ {\bf C 33}, 1129 (1986).

\bibitem{austernpr137}N. Austern, Phys.\ Rev.\ B {\bf 137}, 752 (1965).
\bibitem{km90}N. Keeley and R.S. Mackintosh, Phys.\ Rev.\, C{\bf  90}, 044602 (2014).
\bibitem{remark}R.S. Mackintosh and N. Keeley, Phys.\ Rev.\, C{\bf 98}, 024624 (2018).
\bibitem{tobepub}N. Keeley and R.S. Mackintosh, to be published.

\bibitem{bb63}B. Buck, Phys.\ Rev.\ {\bf 130}, 712 (1963).
\bibitem{fgp63}F.G. Perey, Phys.\ Rev.\ {\bf 131}, 745 (1963).

\bibitem{PL44B}R.S. Mackintosh, Phys.\  Lett.\  {\bf B 44}, 437 (1973).
\bibitem{NPA230}R.S. Mackintosh, Nucl.\ Phys.\ {\bf A 230}, 195 (1974).
\bibitem{kobmac76}A.M. Kobos and R.S. Mackintosh, Phys.\  Lett.\  {\bf B 62}, 127 (1976).
\bibitem{mackob79}R.S. Mackintosh and A.M. Kobos,  J. Phys. G: Nucl. Part. Phys. {\bf 5}, 359 (1979).

\bibitem{OS} L.W. Owen and G.R. Satchler, Phys.\ Rev.\ Lett.\  {\bf 25}, 1720 (1970).
\bibitem{georgiev} B.Z. Georgiev and R.S. Mackintosh,  Phys.\  Lett.\  {\bf B 73}, 250 (1978).
\bibitem{kim1}B.T. Kim, M.C. Kyum, S.W. Hong, M.H. Park, and T. Udagawa, Comput.\ Phys.\ Commun.\, {\bf 71} 150 (1992).
\bibitem{kim2}B.T. Kim, and T. Udagawa, Phys.\ Rev.\ C {\bf 42}, 1147 (1990).
\bibitem{PB}F.G.  Perey and B. Buck, Nucl.\  Phys.\  {\bf 32}, 353 (1962).

\bibitem{RSMSGC} R.S. Mackintosh and S.G. Cooper, J. Phys. G: Nucl. Part. Phys. {\bf 23}, 565 (1997).
\bibitem{LR}D. Lukaszek and G.H. Rawitscher, Phys.\ Rev.\, C{\bf  50}, 968 (1994).
\bibitem{JandM}C.H. Johnson and C. Mahaux, Phys.\ Rev.\, C{\bf  38}, 2589 (1988).

\bibitem{BayeNPA460}D. Baye, Nucl.\ Phys.\ {\bf A 460}, 581 (1986).
\bibitem{vos79}F.K.Vosniakos, N.E. Davison, W.R. Falk, O. Abou-Zeid, and S.P. Kwan,
Nucl.\ Phys.\ {\bf A 332}, 157 (1979).
\bibitem{MR}F. Michel and G. Reidemeister, Z. Phys. A - Atomic Nuclei {\bf 333}, 331 (1989).

\bibitem{cmzp}S.G. Cooper and R.S. Mackintosh, Zeitschrift f\"ur Physik A{\bf 337}, 357 (1990).
\bibitem{npa742}R.S. Mackintosh, Nucl.\ Phys.\ {\bf A 742}, 3 (2004).
\bibitem{NPA589}R.S. Mackintosh and S.G. Cooper,  Nucl.\ Phys.\ {\bf A 589}, 377 (1995).
\bibitem{NPA592}S.G. Cooper and R.S. Mackintosh,  Nucl.\ Phys.\ {\bf A 592}, 338 (1995).
\bibitem{berlin}R.S.Mackintosh and A.A. Ioannides, in \emph{Advanced Methods in the Analysis of Nuclear
Scattering Data}, Lecture notes in Physics Vol. 236,  (Springer Verlag, Berlin), p  283 (1985).

\bibitem{NMS}M.A. Nagarajan, C. Mahaux, and G.R. Satchler, Phys.\ Rev.\ Lett.\, {\bf 54}, 1136 (1985).
\bibitem{ATMR}S. Ait-Tahar, R.S. Mackintosh, and M.A. Russell,   J. Phys. G: Nucl. Part. Phys. {\bf 21}, 577 (1995).

\bibitem{prc67} R.S. Mackintosh and K. Rusek, Phys.\ Rev.\ C{\bf 67},  034607 (2003).
\bibitem{KMprc83} N. Keeley and R.S. Mackintosh, Phys.\ Rev.\ C{\bf 83},  044608 (2011).
\bibitem{skaza} F. Skaza \emph{et al}, Phys.\ Lett.\  B{\bf 619}, 82 (2005).
\bibitem{he8a} R.S. Mackintosh and N. Keeley, Phys.\ Rev.\ C{\bf 83},  057601 (2011).
\bibitem{be10}  R.S. Mackintosh and N. Keeley, Phys.\ Rev.\ C{\bf 76},  024601 (2007).

\bibitem{pmPRC84}D.Y. Pang and R.S. Mackintosh, Phys.\ Rev.\  {\bf C 84}, 064611 (2011).

\bibitem{GM-D}G. Marqu\'inez-Dur\'an, N. Keeley, K.W. Kemper, R.S. Mackintosh, I. Martel, K. Rusek,
and A.M. S\'anchez-Ben\'itez, Phys.\ Rev.\ C{\bf 95},  024602 (2017).

\bibitem{universal}A.A. Ioannides and R.S. Mackintosh, Phys.\ Lett.\ {\bf B 169}, 113 (1986).
\bibitem{mpPRC86}R.S. Mackintosh and D.Y. Pang, Phys.\ Rev.\  {\bf C 86}, 047602 (2012).
\bibitem{kmb}N. Keeley, R.S. Mackintosh, and C. Beck, Nucl.\ Phys.\ {\bf A834}, 792c (2010).
\bibitem{mpPRC88} R.S. Mackintosh and D.Y. Pang, Phys.\ Rev.\  C{\bf 88}, 014608 (2013).

\bibitem{kmPRC77}N. Keeley and R.S. Mackintosh, Phys.\ Rev.\  C{\bf 77}, 054603 (2008).

\bibitem{ARF}R. Alarcon, J. Rapaport, and R.W. Finlay, Nucl.\ Phys.\ {\bf A 462}, 413 (1987).

\bibitem{ermer}M. Ermer, H. Clement, P. Grabmayr, G.J. Wagner, L. Friedrich,
and E. Huttel, Phys.\ Lett.\ B{\bf 188}, 17 (1987).
\bibitem{ermer2} M. Ermer, H. Clement, G. Holetzke, W. Kabitzke, G. Graw, R. Hertenberger,
H. Kader, F. Merz, and P. Schiemenz, Nucl.\  Phys.\  {\bf A533}, 71 (1991).
\bibitem{kmAnnP}A.M. Kobos and R.S. Mackintosh,  Ann. Phys. (NY) {\bf 123},  296 (1979).


\bibitem{shamu}R.E. Shamu, J. Barnes, S.M. Ferguson. G. Haouat, and J. Lachkat,
J. Phys. G: Nucl. Part. Phys.  {\bf 17}, 525 (1991).
\bibitem{kmr}A.M. Kobos, R.S. Mackintosh, and J.R. Rook, Nucl.\ Phys.\ {\bf A 389}, 205  (1982).
\bibitem{rsm79}R.S. Mackintosh, J. Phys. G: Nucl. Part. Phys.  {\bf 5}, 1587 (1979).
\bibitem{cordero}R.S. Mackintosh and L.A. Cordero-L., Phys. Lett {\bf B 68}, 213 (1977).
\bibitem{kobmac79}A.M. Kobos and R.S. Mackintosh,  J. Phys. G: Nucl. Part. Phys. {\bf 5}, 97 (1979).
\bibitem{kobmac81}A.M. Kobos and R.S. Mackintosh, Acta Physica Polonica,  {\bf B} 12, 1029 (1981).


\bibitem{available}R.S. Mackintosh, unpublished (2016); postscript files available on request.

\bibitem{rm-he3}R.S.Mackintosh, unpublished (1978); scanned copy available on request.
\bibitem{brum-he3}Y.-W. Lui, O. Karban, S. Roman, R.K. Bhowmik, J.M. Nelson, and E.C. Pollacco,
Nucl.\ Phys.\ {\bf A 333}, 205  (1980).
\bibitem{cage}M.E. Cage, D.L. Clough, A.J. Cole, J.B.A England, G.J. Pyle, P.M. Rolph,
L.H. Watson, and D.H. Worledge,  Nucl.\ Phys.\ {\bf A 183}, 449  (1972).

\bibitem{preprint}R.S. Mackintosh, unpublished preprint, available.
\bibitem{ttPRC4}D. R. Thompson and Y. C. Tang,  Phys.\ Rev.\, C{\bf 4}, 306 (1971).
\bibitem{ttbPRC5}D. R. Thompson, Y. C. Tang, and R.E. Brown,  Phys.\ Rev.\, C{\bf 5}, 1939 (1972).
\bibitem{gmttPRC6}G. W. Greenlees, W. Makofske, Y. C. Tang, and D. R. Thompson,  Phys.\ Rev.\, C{\bf 6}, 2057 (1972).

\bibitem{PRC54}S.G. Cooper and R.S. Mackintosh, Phys.\ Rev.\ C{\bf 54},  3133 (1996).
\bibitem{NPA618}S.G. Cooper, Nucl.\ Phys.\ {\bf A 618}, 87 (1997).
\bibitem{npa713}H. Heiberg-Andersen, R.S. Mackintosh, and J.S. Vaagen, Nucl.\
Phys.\ {\bf A 713}, 63 (2003).
\bibitem{NPA517}S.G. Cooper and R.S. Mackintosh,  Nucl.\ Phys.\ {\bf A 517}, 285 (1990).
\bibitem{kmprc97}N. Keeley and R.S. Mackintosh, Phys.\ Rev.\, C{\bf  97}, 014605 (2018).

\bibitem{eegross}E.E. Gross, R.H. Bassel, L.N. Blumberg, B.J. Morton, A. van der Woude,
and A. Zucker, Nucl.\ Phys.\ {\bf A 102}, 673 (1967).

\bibitem{HJM}D.J. Horen, C.H. Johnson, and A.D. MacKellar, Phys. Lett. {\bf B 161}, 217 (1085).
\bibitem{HJFMC} D.J. Horen, C.H. Johnson, J.L. Fowler, A.D. MacKellar, and B. Castel,
Phys.\ Rev.\ C{\bf 34}, 429 (1986).
\bibitem{JHM}C.H. Johnson,  D.J. Horen, and C. Mahaux,  Phys.\ Rev.\ C{\bf 36}, 2252 (1987).
\bibitem{JW}C.H. Johnson and R.R. Winters,   Phys.\ Rev.\ C{\bf 37}, 2340 (1988).
\bibitem{JJM}J.-P. Jeukenne, C.H. Johnson,  and C. Mahaux,       Phys.\ Rev.\ C{\bf 38}, 2573 (1988).


\bibitem{WH}T. Wada and H. Horiuchi, Progress of Theoretical Physics, {\bf 80},
488 (1988); {\bf 80}, 502 (1988).
\bibitem{horiuchi}H. Horiuchi, \emph{Proc. Int. Conf. on Clustering Aspects of Nuclear  Structure
 and Nuclear Reactions} (Chester, 1984), ed. J.S. Lilley and M.A. Nagarajan (Reidel, Dordrecht), p. 35.

\bibitem{ATMCW}S. Ait-Tahar, R.S. Mackintosh,  S.G. Cooper, and T. Wada, Nucl.
Phys. {\bf A562}, 101 (1993).

\bibitem{KRS}Y. Kondo, B.A. Robson, and R. Smith, Phys.\  Lett.\  {\bf B 227}, 310  (1989).
\bibitem{ATCM}S. Ait-Tahar,  S.G. Cooper, R.S. Mackintosh, Nucl. Phys. {\bf A542}, 499 (1992).
\bibitem{cerr70} R.A. Chatwin, J.S. Eck, D. Robson, and A. Richter,  Phys.\ Rev.\  C{\bf 1}, 795 (1970).
\bibitem{bisson1}A.E. Bisson and R.H. Davis, Phys.\ Rev.\  Lett\ {\bf 22}, 542 (1969).
\bibitem{bisson2}A.E. Bisson, K.A. Eberhard, and R.H. Davis, Phys.\ Rev.\ C{\bf 1}, 539 (1970).


\bibitem{gaohe}C. Gao and G. He, Phys.\  Lett.\  {\bf B 282}, 16  (1992).
\bibitem{ATMC}S. Ait-Tahar, R.S. Mackintosh,  S.G. Cooper,   Nucl. Phys. {\bf A561}, 285 (1993).

\bibitem{bohn}H. Bohn, K.A. Eberhart, R. Vandenbosch, K.G. Bernhardt, R. Bangert, and Y-d. Chan,
Phys.\ Rev.\ C{\bf 16}, 665 (1977).

\bibitem{trcka}D.E. Trcka, A.D. Frawley, K.W. Kemper, D. Robson, J.D. Fox, and E.G. Myers,
Phys.\ Rev.\ C{\bf 41}, 2134 (1990).

\bibitem{Retal}E.L. Reber, K.W. Kemper, P.V. Green, P.L. Kerr, A.J. Mendez, E.G. Myers,
and B.G. Schmidt, Phys.\ Rev.\ C{\bf  49}, R1 (1994).

%\bibitem{ATMR}S. Ait-Tahar, R.S. Mackintosh, and M.A. Russell,   J. Phys. G: Nucl. Part. Phys. {\bf 21}, 577 (1995).

\bibitem{3author} R.S. Mackintosh, Y. Hirabayashi, and S. Ohkubo, Phys.\ Rev.\ {\bf C 91}, 024616 (2015).
\bibitem{OH1}S. Ohkubo and Y. Hirabayashi,  Phys.\ Rev.\ {\bf C 89}, 051601 (2014).
\bibitem{OH2}S. Ohkubo and Y. Hirabayashi,  Phys.\ Rev.\ {\bf C 89}, 061601 (2014).

\bibitem{116}R.S. Mackintosh, Phys.\ Rev.\ {\bf C 94}, 034602 (2016).

\bibitem{lee}S.Y. Lee, Nucl. Phys. {\bf A311}, 518 (1978).
\bibitem{MKsplines} R.S. Mackintosh and A.M. Kobos, Phys.\  Lett.\  {\bf B 92}, 59  (1980).
\bibitem{KSM}A.M. Kobos, G.R. Satchler,  and R.S. Mackintosh, Nucl. Phys. {\bf A395}, 248 (1983).
\bibitem{KMPRC26}A.M. Kobos and R.S. Mackintosh, Phys.\ Rev.\ {\bf C 26}, 1766 (1982).

%\bibitem{pang1}D.Y. Pang, P. Roussel-Chomaz, H. Savajols, R.L Varner, and R. Wolski,
%Phys.\ Rev.\ C{\bf 79}, 024615 (2009).
\bibitem{pang2}D.Y. Pang, Y.L. Ye, and F.R. Xu, Phys.\ Rev.\ C{\bf 83}, 064619
(2011).
\bibitem{OHO}S. Ohkubo, Y. Hirabayashi, and A.A. Ogloblin,  Phys.\ Rev.\ {\bf C 92}, 051601 (2015).

\bibitem{nunes}F.M. Nunes, A. Lovell, A. Ross, L.J. Titus, R.J. Charity, W.H. Dickhoff, M.H. Mahzoon,
J. Sarich, and S.M. Wild, arXiv:1509.047001




\end{thebibliography}
\end{document}